%% file: TGA.tex
\documentclass{elsart}

\usepackage{amsmath,amsfonts,amssymb,graphicx,multirow}
\usepackage{algorithm,algorithmic}
\usepackage[font=footnotesize]{caption}
\usepackage[font=footnotesize]{subcaption}
\usepackage{color}
\usepackage{booktabs}
\usepackage{url}

\begin{document}

\begin{frontmatter}

\title{Speeding up Local Optimization in Vehicle Routing with Tensor-based GPU Acceleration}

\author[LERIA]{Zhenyu Lei},
\ead{zhenyu.lei@etud.univ-angers.fr}
\author[LERIA]{Jin-Kao Hao}
\ead{jin-kao.hao@univ-angers.fr}
\address[LERIA]{LERIA, Universit$\acute{e}$ d'Angers, 2 Boulevard Lavoisier, 49045 Angers, France}
\author[HUST]{Qinghua Wu}
\ead{qinghuawu1005@gmail.com}
\address[HUST]{School of Management, Huazhong University of Science and Technology, No.1037, Luoyu Road, Wuhan, China}

\maketitle

\begin{abstract}
Local search is central to many high-performing metaheuristic algorithms for the vehicle routing problem (VRP) and its variants. However, neighborhood exploration is computationally expensive, especially for large or highly constrained instances. This paper addresses this challenge by proposing an original Tensor-based GPU Acceleration (TGA) framework with two implementations. TGA accelerates move evaluation for common local search operators through an attribute-based solution tensor representation and fully tensorized operator implementations that exploit GPU parallelism. The framework is extensible to a wide range of VRP variants and can be seamlessly integrated into local search-based algorithms. Extensive experiments on three routing problems demonstrate significant advantages over traditional CPU-based implementations. A detailed analysis further reveals performance characteristics, potential bottlenecks, and directions for future improvement.

\noindent \emph{Keywords}: Tensor-based GPU acceleration; Local search; Vehicle routing.
\end{abstract}

\end{frontmatter}

\section{Introduction}
\label{sec:introduction}
The Vehicle Routing Problem (VRP) \cite{dantzig1959truck,clarke1964scheduling} is one of the most well-known combinatorial optimization problems and is very computationally challenging due to its NP-hard nature \cite{lenstra1978complexity}. The VRP involves dispatching a fleet of vehicles to serve a set of customers, with each route starting and ending at a depot, while satisfying certain constraints and minimizing the total operational cost. As a fundamental problem in combinatorial optimization, the VRP finds numerous real-world applications in the fields of transportation, logistics, supply chain management, and so on. 

Since its introduction, the VRP and its variants have been extensively studied. Methods for solving these problems include exact, approximate, heuristic, and metaheuristic approaches. Due to the high complexity, exact and approximate methods \cite{laporte1992vehicle,andelmin2017exact,praxedes2024unified} are typically limited to small and medium instances, while heuristics and metaheuristics are used for larger instances to obtain good suboptimal solutions efficiently. These include simulated annealing \cite{chiang1996simulated,rodriguez2024new}, tabu search \cite{toth2003granular,gmira2021tabu}, genetic algorithms \cite{baker2003genetic,tasan2012genetic}, and memetic algorithms \cite{cattaruzza2014memetic,vidal2020hybrid}, among others. Comprehensive literature reviews on the classification of VRPs and the existing methods can be found in surveys \cite{laporte2009fifty,vidal2013heuristics,braekers2016vehicle,tan2021vehicle}. As VRPs grow in size and complexity, computational efficiency remains a challenge, especially with complex constraints. Local search, a key component of high-performing solvers, relies on iterative neighborhood exploration, which is computationally intensive for large or constrained instances. Improving the efficiency of neighborhood evaluation is thus critical for enhancing search performance.

With recent advances in computer hardware, GPU (Graphics Processing Unit) has become increasingly promising for accelerating computationally intensive tasks in combinatorial optimization. With their highly parallel architecture and thousands of cores, GPUs are well suited for large-scale data processing and tasks requiring extensive parallelism, offering significant speedups over traditional CPU-based approaches for solving complex optimization problems. Several studies have explored redesigning classical heuristics and metaheuristics to exploit GPU parallelism. \cite{rashid2020efficient} implemented hill climbing and simulated annealing on GPUs and reported substantial speedups, while \cite{cheng2019accelerating} surveyed GPU-based genetic algorithms across different parallel models and granularities. \cite{goudet2023massively,goudet2024large} proposed parallel hybrid evolutionary algorithms that exploit algorithmic-level parallelism by running multiple local search processes in parallel within a large population on GPU. \cite{NOGUEIRA2024} developed a GPU-based tabu search for the maximum diversity problem, achieving significant acceleration in neighborhood evaluation.

Recent studies have also explored GPU-based approaches for routing problems, which can be categorized into coarse-grained parallelism, accelerating high-level algorithmic components, and fine-grained parallelism, speeding up neighborhood evaluation in local search. \cite{abdelatti2020improved} proposed a hybrid genetic algorithm with 2-opt local search for the Capacitated VRP (CVRP) on GPU, parallelizing all algorithmic components. \cite{zhang2019gmma} developed a GPU-based multi-objective memetic algorithm for VRPs with route balancing, employing solution-level parallelism and route-level parallelism, though they did not exploit finer node-level parallelism. \cite{schulz2013efficient} achieved significant speedups by parallelizing 2-opt and 3-opt operators for the CVRP. \cite{yelmewad2020gpu} extended GPU-based local search with or-opt, swap, and relocate operators, improving performance but limited to basic travel distance evaluation. \cite{coelho2016integrated} proposed a GPU-based variable neighborhood search for VRPs with deliveries and selective pickups, incorporating multiple local search operators, yet primarily optimized travel distance and struggled with complex constraints.

These studies demonstrate the potential of GPU acceleration for solving VRPs, but several limitations remain. Most methods focus on basic VRP variants or simple operators, with limited support for complex constraints and diverse problem settings. Moreover, current approaches rely heavily on low-level CUDA implementations. While powerful, CUDA poses a high barrier to entry due to its steep learning curve and the need for manual management of data transfers, memory layouts, parallel execution, and thread synchronization. This makes widespread adoption challenging, particularly for researchers in operations research without specialized GPU expertise.

To address these gaps, we propose a fine-grained Tensor-based GPU Acceleration (TGA) framework to improve the computational efficiency of local search-based algorithms for complex VRPs. The main contributions are: (1) The TGA framework, along with two implementations, is proposed to accelerate move evaluation by tensorizing local search operators based on an attribute-based solution tensor representation, thereby fully leveraging GPU parallelism. (2) TGA is highly extensible, supporting a wide range of VRP variants and integration into different local search solvers. (3) Unlike CUDA-based approaches, TGA can be implemented using high-level tensor libraries such as PyTorch, substantially lowering the barrier to GPU acceleration for operations research researchers. (4) In-depth theoretical and experimental analyses provide insights into performance characteristics and potential bottlenecks, offering guidance for future research. 

To demonstrate its performance and extensibility, we apply TGA to three representative VRPs: CVRP, VRP with Time Windows (VRPTW), and VRP with Simultaneous Pickup and Delivery and Time Windows (VRPSPDTW). Results on benchmark instances confirm that the TGA effectively handles diverse constraints and achieves substantial efficiency gains over traditional CPU-based implementations.

The remainder of this paper is organized as follows. Section~\ref{sec:problem} formally defines the VRPs. Section~\ref{sec:gpu} presents an overview of GPU architecture and tensor computation, forming the basis of the proposed framework. Section~\ref{sec:move_eval} introduces essential background on attribute matrices and sequence concatenation-based move evaluation. Section~\ref{sec:method} presents the tensor-based GPU acceleration framework. Section~\ref{sec:computational_results} reports computational results on benchmark instances and provides an in-depth analysis of the framework's advantages and limitations. Section~\ref{sec:discussion} discusses its extensibility and limitations, while Section~\ref{sec:conclusion} concludes the paper. A supplementary document is provided that summarizes the key notations (Section~A1), describes the algorithmic framework of the tested algorithms (Section~A2), and reports detailed computational results (Sections~A3-A4).

\section{Problem definition}
\label{sec:problem}
This section introduces the three typical routing problems that serve as the case studies: the Capacitated VRP, the VRP with Time Windows, and the VRP with Simultaneous Pickup and Delivery and Time Windows. The VRPSPDTW is a generalization of both the CVRP and the VRPTW. Therefore, we will focus on defining the VRPSPDTW, while showing how it is reduced to these more specific problems.

All three problems can be defined on a complete graph $\mathcal{G}=(\mathcal{V}, \mathcal{E})$, where the vertices or nodes $\mathcal{V} = \{v_0, v_1, \dots, v_{N_C}\}$ represent the depot  $v_0$ and the $N_C$ customers $\{v_1, \dots, v_{N_C}\}$. The edges $\mathcal{E} = \{e_{ij} | v_i, v_j \in \mathcal{V}\}$ represent the connections between these nodes. Each customer node $v_i \in \mathcal{V}$ ($i \neq 0$) is associated with a delivery demand $d_i$ and a pickup demand $p_i$. This means that the vehicle must deliver $d_i$ units of goods from the depot $v_0$ to $v_i$ and pick up $p_i$ units from $v_i$ back to the depot. Furthermore, each node $v_i$ has a time window [$e_i$, $l_i$] and a service time $s_i$, which specifies the earliest and latest time for the service, as well as the time spent at node $v_i$ for the service. The depot has a time window [$e_0$, $l_0$], specifying the earliest departure and latest return time, with its service time $s_0$ set to 0. Additionally, the travel distance and travel time matrices, $\mathcal{C} = (c_{ij})$ and $\mathcal{T} = (t_{ij})$, store the travel distance and time for each edge $e_{ij} \in \mathcal{E}$. For the VRPTW and the CVRP, the pickup demand $p_i = 0$. Additionally, the CVRP does not consider time-related attributes such as travel time, service time, or time windows. 

A fleet of $M$ homogeneous vehicles, each with a capacity $\mathcal{Q}$, serves all customers, starting and ending at the depot $v_0$, while satisfying various constraints, including capacity and time windows. A solution $S$ is represented as a set of closed routes $S = \{R_1, \dots, R_M\}$, where each route $R_i$ consists of a sequence of nodes $\{n_{i,0}, n_{i,1}, \dots, n_{i,\mathcal{L}_i}, n_{i, \mathcal{L}_i+1}\}$ visited by the $i$-th vehicle. Here, $n_{i,j}$ denotes the $j$-th node in route $R_i$ and $\mathcal{L}_i$  is the number of served customers in that route. Thus, the length of route $R_i$ is $|R_i|=\mathcal{L}_i+2$. Note that both the first and last nodes in route $R_i$ are the depot $v_0$ (i.e., $n_{i,0} = v_0$ and $n_{i,\mathcal{L}_i+1} = v_0$). To ensure that the load of each vehicle does not exceed its capacity $\mathcal{Q}$, the condition $q_{n_{i,j}} \leq \mathcal{Q}$ must hold for all $n_{i,j} \in R_i$, where $q_{n_{i,j}}$ represents the load of the $i$-th vehicle after visiting node $n_{i,j}$. For variants that incorporate time windows, let $a_{n_{i,j}}$ denote the arrival time at node $n_{i,j}$. Arriving before $e_{n_{i,j}}$ results in a wait time of  $w_{n_{i,j}} = \max\{e_{n_{i,j}} - a_{n_{i,j}}, 0\}$, while arriving after $l_{n_{i,j}}$ renders the route infeasible.
\begin{align}
\label{eq:objective}
\mathrm{Minimize} \quad f(S) &= \mu_1 \cdot M + \mu_2 \cdot D(S)  \\
\mathrm{Subject \ to} \quad D(S) &= \sum_{i=1}^{M} \sum_{j=0}^{\mathcal{L}_i} c_{n_{i,j}n_{i,j+1}} \nonumber \\
S &= \{R_1, \dots, R_M\} \nonumber \\
R_i &= \{n_{i,0}, n_{i,1}, \dots, n_{i,\mathcal{L}_i}, n_{i,\mathcal{L}_i+1}\}, \quad i=1,\dots,M \nonumber
\end{align}
Typically, we can describe the objective function of these problems as minimizing the total cost, which is the weighted sum of the number of vehicles and the travel distance with predefined weights, as shown in Equation~(\ref{eq:objective}), where $\mu_1$ and $\mu_2$ represent the costs or weights assigned to the vehicle dispatching and the travel distance, respectively.

\section{GPU and tensor computation}
\label{sec:gpu}
This section provides background on GPU architecture, CUDA programming, and tensor computation, forming the basis of our acceleration method.
\subsection{GPU architecture}
\label{sec:gpu_architecture}
Originally designed for graphics rendering, GPUs are specialized parallel processors that are now widely used for general-purpose computing in scientific simulations, data analytics, and artificial intelligence. Unlike CPUs, which prioritize low-latency execution of sequential instructions through complex control logic and extensive caching, GPUs dedicate most of resources to arithmetic units, enabling thousands of lightweight threads to execute concurrently.

NVIDIA GPUs dominate high-performance and scientific computing due to their performance and mature software ecosystem. The fundamental unit of NVIDIA GPUs, the \emph{Streaming Multiprocessor} (SM), integrates numerous CUDA cores, registers, shared memory, and caches, enabling thousands of threads to cooperate efficiently. Execution on NVIDIA GPUs follows the SIMT (Single Instruction, Multiple Threads) model, where threads are organized into 32-thread \emph{warps} that execute the same instruction on different data, simplifying scheduling and enabling scalable fine-grained parallelism. Additionally, the GPU memory hierarchy supports high-throughput workloads, with large high-latency global memory, fast per-SM shared memory for inter-thread communication, and low-latency per-thread registers.
\begin{table}[!ht]
\centering
\caption{Comparison of NVIDIA GPUs: V100, A100, H100.}
\label{tab:gpu_comparison}
\resizebox{\textwidth}{!}{%
\begin{tabular}{lllcccccc}
\hline
GPU & Year & Arch. & SMs & Cores/SM & Total Cores & Memory & Memory Bandwidth & Compute Throughput (TFLOPS) \\
\hline
V100 & 2017 & Volta  & 80  & 64   & 5,120  & 16/32 GB HBM2  & $\sim$900-1,134 GB/s & $\sim$15.7 (FP32) / $\sim$7.8 (FP64) \\
A100 & 2020 & Ampere & 108 & 64   & 6,912  & 40/80 GB HBM2e & $\sim$1.6-2.0 TB/s   & $\sim$19.5 (FP32) / $\sim$9.7 (FP64) \\
H100 & 2022 & Hopper & 132 & 128  & 16,896 & 80 GB HBM3     & $\sim$3.0-3.2 TB/s   & $\sim$67 (FP32) / $\sim$34 (FP64) \\
\hline
\end{tabular}%
}
\end{table}

Table~\ref{tab:gpu_comparison} summarizes three recent NVIDIA GPU generations, showing consistent increases in SMs, CUDA cores, memory capacity, memory bandwidth, and floating-point throughput. More SMs, cores, and higher compute throughput enhance parallelism and speed up compute-intensive workloads. Larger and faster memory supports larger datasets locally, while higher memory bandwidth enables faster data transfer between memory and compute units. Together, these improvements substantially boost modern GPUs' ability to handle large-scale, complex parallel workloads in scientific and industrial applications.

\subsection{CUDA programming}
\label{sec:cuda}
NVIDIA's Compute Unified Device Architecture (CUDA) is the primary framework for general-purpose GPU programming on NVIDIA hardware. It provides a C/C++ based model in which developers write kernels executed in parallel by many lightweight threads. CUDA exposes hardware concepts such as threads, blocks, warps, and multiple levels of memory, offering fine-grained control over parallel execution and data movement.

Despite its flexibility, CUDA development is challenging. Programmers must explicitly manage device memory, data transfers between CPU and GPU, and thread synchronization. High performance requires deep knowledge of GPU architecture, including how to design coalesced memory accesses, avoid bank conflicts, minimize warp divergence, and tune occupancy. CUDA programming is thus inherently more time-consuming and error-prone than high-level CPU programming, limiting adoption among operations researchers who primarily focus on algorithm design or problem modeling and often lack these specialized skills.

\subsection{Tensor computation}
\label{sec:tensor_computing}
In scientific computing and machine learning, tensor computation has emerged as an effective paradigm for exploiting GPU acceleration. A \emph{tensor} is a multi-dimensional array that generalizes scalars (0-D), vectors (1-D), and matrices (2-D) to higher dimensions. On GPUs, tensors are typically stored as contiguous memory blocks, enabling coalesced memory access and efficient parallel execution. Tensor operations are inherently parallel, as computations on different elements can be distributed across thousands of GPU cores, allowing tensor-based workloads to fully exploit the massive parallelism of modern GPU architectures.

High-level frameworks such as PyTorch and TensorFlow provide a rich set of tensor operations, including element-wise arithmetic, slicing, reshaping, and reduction. These frameworks automatically translate tensor operations into highly optimized GPU kernels, relieving developers from explicitly managing threads, memory transfers, and kernel launches, which are required in low-level CUDA programming. By abstracting away hardware-specific details, tensor computation enables researchers to concentrate on algorithmic design while still achieving high-performance execution. 

Thus, in this study, we model the solution with tensor representation, and implement move evaluation of local search operators using tensor operations. This design enables efficient exploitation of GPU parallelism through high-level tensor frameworks, making GPU acceleration more accessible and practical for researchers in operations research.
\section{Move evaluation and attribute matrix}
\label{sec:move_eval}
Efficient move evaluation is a critical step in the search process of local search algorithms and directly affects both solution quality and search efficiency. Our proposed tensor-based GPU acceleration (TGA) framework builds upon a constant-time move evaluation method \cite{vidal2014unified,liu2021memetic,lei2024memetic}. Although this method has not been explicitly named in prior works, we refer to it in this paper as \emph{sequence concatenation-based move evaluation}.
\subsection{Sequence concatenation-based move evaluation}
\label{sec:sequence_concatenation}
Sequence concatenation-based move evaluation conceptualizes common local search move operators, such as \emph{Relocate}, \emph{Swap}, \emph{2-opt*}, and \emph{2-opt}, as the concatenation of multiple route subsequences. This allows for efficient neighborhood evaluations with respect to both the objective function and the associated constraints.
\begin{figure}[!htbp]
    \centering
    \begin{subfigure}{0.25\textwidth}
        \includegraphics[width=\textwidth]{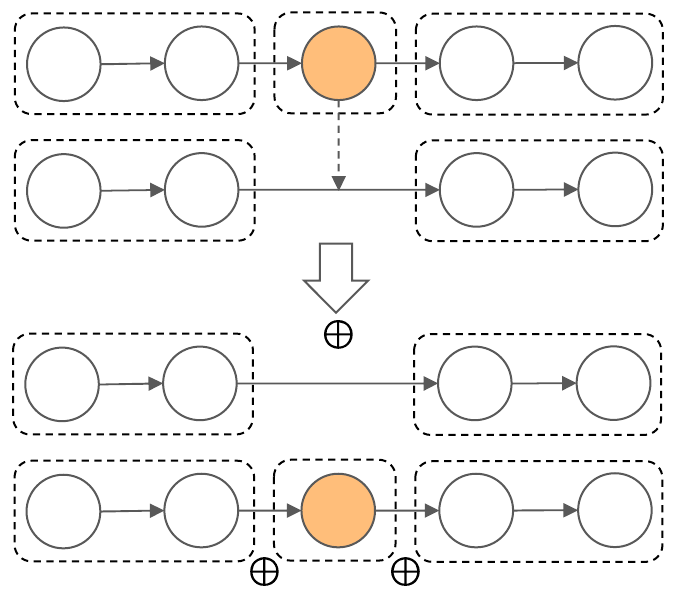}
        \caption{\emph{Inter-Relocate}}
        \label{fig:o-inter-relocate}
    \end{subfigure}
    \hfill
    \begin{subfigure}{0.25\textwidth}
        \includegraphics[width=\textwidth]{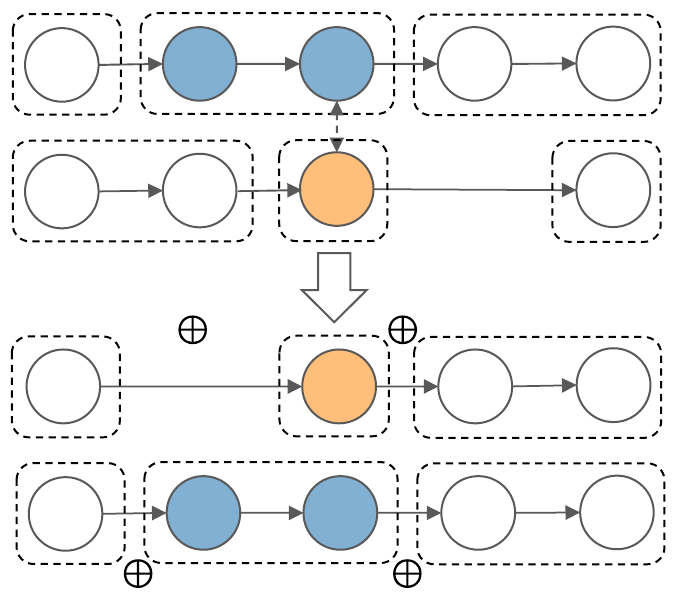}
        \caption{\emph{Inter-Swap}}
        \label{fig:o-inter-swap}
    \end{subfigure}
    \hfill
    \begin{subfigure}{0.25\textwidth}
        \includegraphics[width=\textwidth]{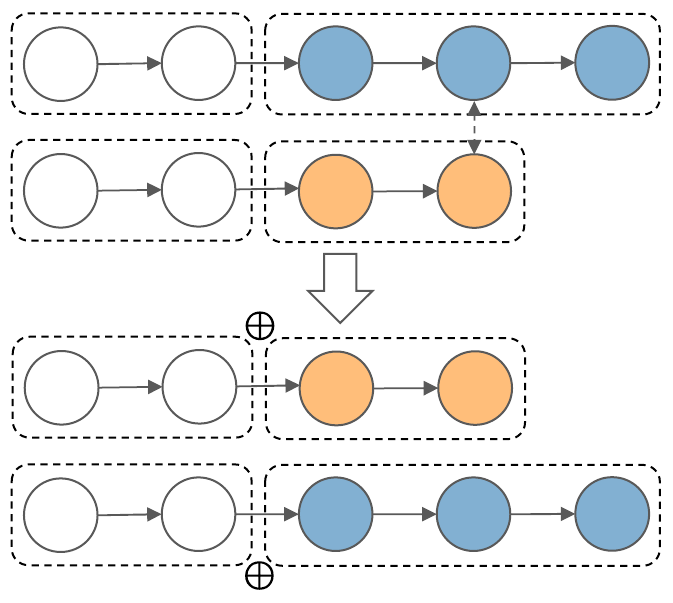}
        \caption{\emph{2-opt*}}
        \label{fig:o-2opt-star}
    \end{subfigure}
    \hfill
    \begin{subfigure}{0.25\textwidth}
        \includegraphics[width=\textwidth]{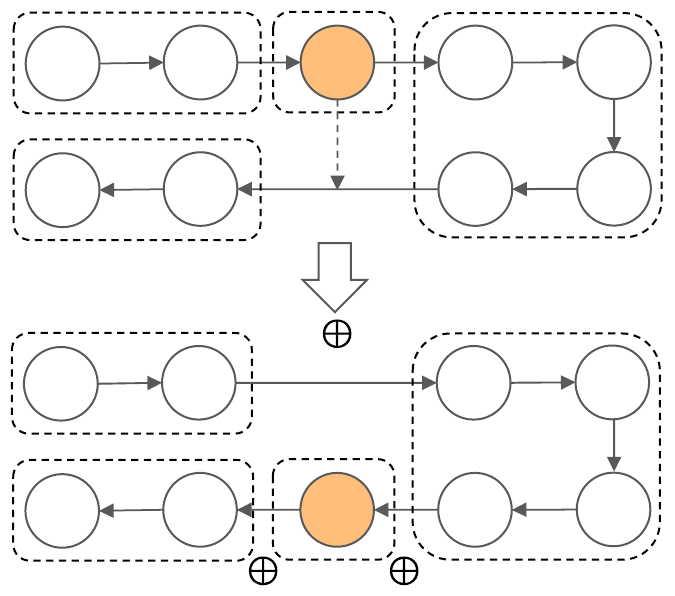}
        \caption{\emph{Intra-Relocate}}
        \label{fig:o-intra-relocate}
    \end{subfigure}
    \hfill
    \begin{subfigure}{0.25\textwidth}
        \includegraphics[width=\textwidth]{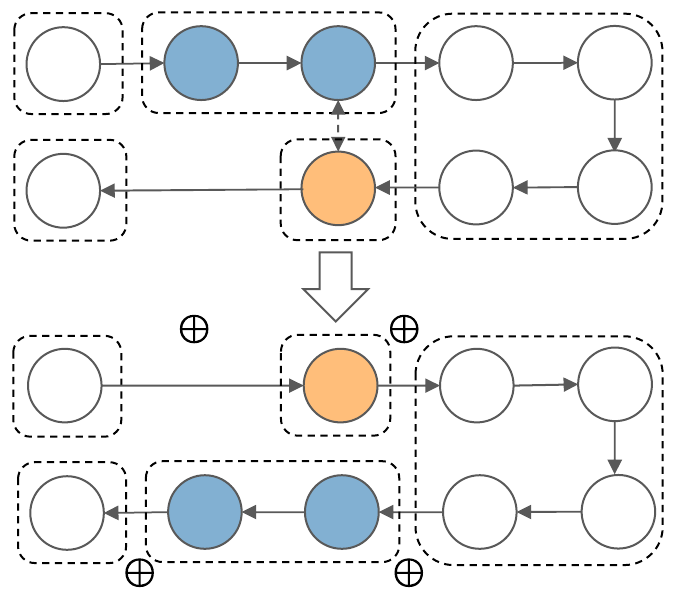}
        \caption{\emph{Intra-Swap}}
        \label{fig:o-intra-swap}
    \end{subfigure}
    \hfill
    \begin{subfigure}{0.25\textwidth}
        \includegraphics[width=\textwidth]{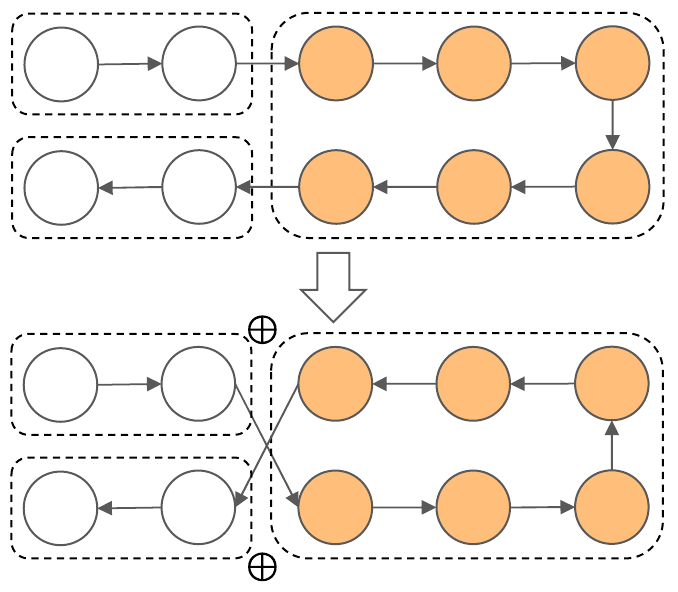}
        \caption{\emph{2-opt}}
        \label{fig:o-2opt}
    \end{subfigure}
\caption{Illustration of operators and their concatenation operations. Dotted boxes represent the route subsequences. Symbols $\oplus$ denote the concatenation operations.}
\label{fig:operators}
\end{figure}

The \emph{Relocate} operator moves a sequence of nodes to a new position within the same route or a different route, while the \emph{Swap} operator exchanges two sequences of nodes. The \emph{2-opt*} operator, applied between two routes, breaks one edge in each route, exchanges the subsequences, and reconnects them to form new routes. In contrast, the \emph{2-opt} move operates within a single route by breaking two non-adjacent edges and reconnecting the reversed subsequence. Note that \emph{2-opt} is only suitable for problems with symmetric distance matrices and without directed constraints, such as time windows. Based on whether the operators act within a single route or between two routes,
these operators can be classified into two categories: \emph{intra-route} operators, which act within a single route (\emph{Intra-Relocate}, \emph{Intra-Swap} and \emph{2-opt}), and \emph{inter-route} operators, which operate between two routes (\emph{Inter-Relocate}, \emph{Inter-Swap}, and \emph{2-opt*}). Figure~\ref{fig:operators} illustrates the concatenation operations for each operator. The concatenation operations can also be extended to other routing operators with similar characteristics, such as 3-opt, Or-opt, and Cross-exchange.

\subsection{Attribute matrices}
\label{sec:attribute_mat} 
\begin{figure}[!htbp]
    \centering
    \begin{subfigure}{0.45\textwidth}
        \includegraphics[trim=5 5 5 5,
    clip,width=\textwidth]{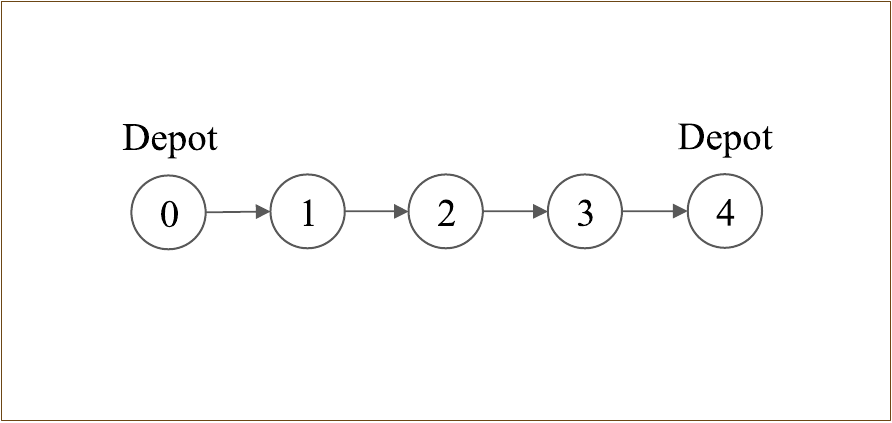}
        \caption{Route representation.}
        \label{fig:route_rep}
    \end{subfigure}
    \hfill
    \begin{subfigure}{0.5\textwidth}
        \includegraphics[trim=5 5 5 5,
    clip,width=\textwidth]{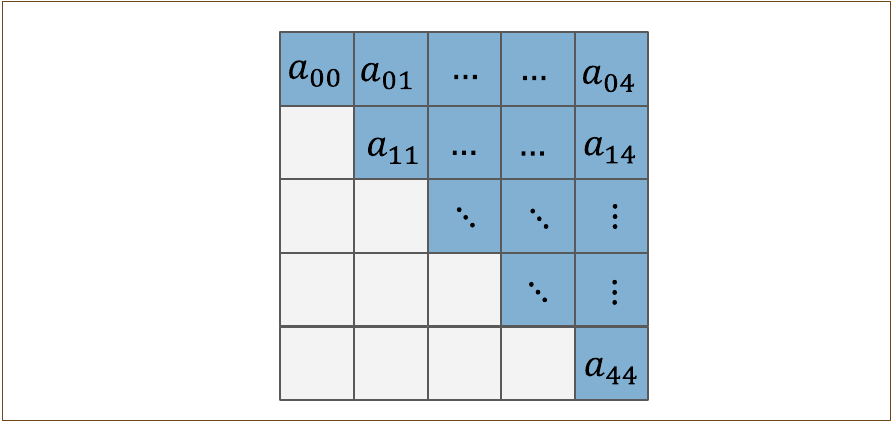}
        \caption{Attribute matrix}
        \label{fig:attr_mat}
    \end{subfigure}
\caption{Illustration of a route of length $|R|=5$ and its corresponding attribute matrix. The first~(0) and last~(4) nodes represent the depot.}
\label{fig:attribute}
\end{figure}

To enable efficient sequence concatenation-based move evaluation, we introduce \emph{attribute matrices}, which store the attribute values of all feasible route subsequences. These values are precomputed during the update stage, enabling constant-time computation of concatenated sequences in the move evaluation stage. Specifically, for each route and for each relevant attribute, an \emph{attribute matrix} is constructed to record the attribute values of all feasible subsequences. Figure~\ref{fig:attribute} illustrates a route of length $|R|=5$ along with its attribute matrix, where $a_{ij}$ denotes the attribute value associated with the subsequence from the $i$-th node to the $j$-th node of the route. For example, for the travel distance attribute, $a_{ij}$ corresponds to the cumulative travel distance between the $i$-th and $j$-th nodes. Note that only the upper triangular portion of the matrix contains valid entries. These attributes are tailored to specific problem settings and constraints, providing strong adaptability across various VRP variants. 

Building upon these attribute matrices, constant-time move evaluation is achieved through sequence concatenation operations, which constitute a core component of the proposed tensor-based GPU acceleration framework.

\subsection{Calculations of concatenation operations}
\label{sec:calculation_concat}
This section details the calculations of concatenation operations using attribute matrices. We focus on commonly used attributes: \emph{travel distance}, \emph{load related attributes}, and \emph{time window related attributes}, which are sufficient for the studied VRPs (CVRP, VRPTW, and VRPSPDTW) and widely adopted in vehicle routing. This concatenation operation can be extended to attributes required by other VRP variants, such as cumulative distance, load-dependent cost as discussed in \cite{vidal2014unified}.

Let $\sigma$ denote a route subsequence. A single-node subsequence is denoted by $\sigma^{0} = \{v_k\}$, while $\sigma_1 = \{v_i, \dots, v_j\}$ and $\sigma_2 = \{v_k, \dots, v_l\}$ represent two generic subsequences. The concatenation of $\sigma_1$ and $\sigma_2$ is denoted by $\sigma_1 \oplus \sigma_2$. The following subsections describe how attributes are updated under concatenation.

\subsubsection{Travel distance}
Travel distance is a fundamental attribute in routing problems. Let $D(\sigma)$ denote the travel distance of a subsequence $\sigma$. For a single-node subsequence $\sigma^{0}$ and the concatenation of two route subsequences $\sigma_1$ and $\sigma_2$, the travel distance is computed as follows. 
\begin{subequations}
\begin{align}
& D(\sigma^0) = 0 \\
& D(\sigma_1 \oplus \sigma_2) = D(\sigma_1) + c_{jk} + D(\sigma_2) \label{eq:dist}
\end{align}
\end{subequations}
where $c_{jk}$ denotes the travel distance between the last node $v_j$ of $\sigma_1$ and the first node $v_k$ of $\sigma_2$.
\subsubsection{Load and capacity}
To evaluate load and capacity constraints, we maintain three load-related attributes for each route subsequence $\sigma$: the incoming load $L_I(\sigma)$, the outgoing load $L_O(\sigma)$, and the maximum load $L_M(\sigma)$. The computation of these attributes is defined as follows. 
\begin{subequations}
\begin{align}
&L_I(\sigma^{0}) = d_k, \quad
L_O(\sigma^{0}) = p_k, \quad
L_M(\sigma^{0}) = \max\{d_k, p_k\} \\
&L_I(\sigma_1 \oplus \sigma_2) = L_I(\sigma_1) + L_I(\sigma_2) \\
&L_O(\sigma_1 \oplus \sigma_2) = L_O(\sigma_1) + L_O(\sigma_2) \\
&L_M(\sigma_1 \oplus \sigma_2) = \max\{L_{M}(\sigma_1) + L_{I}(\sigma_2), L_{O}(\sigma_1) + L_{M}(\sigma_2)\}
\end{align}
\end{subequations}
For CVRP and VRPTW, where pickup demands are zero ($p_k = 0$), the formulation simplifies and only the maximum load needs to be maintained:
\begin{subequations}
\begin{align}
&L_M(\sigma^{0}) = d_k \\
&L_M(\sigma_1 \oplus \sigma_2) = L_M(\sigma_1) + L_M(\sigma_2)
\end{align}
\end{subequations}
\subsubsection{Time windows}
Time window constraints can be handled using the time-warp technique \cite{nagata2010penalty,vidal2013hybrid}, which allows the evaluation of time window violations within subsequences. Each subsequence $\sigma$ maintains the earliest and latest arrival times ($T_E(\sigma)$ and $T_L(\sigma)$), duration $T_D(\sigma)$, and time window violation $T_V(\sigma)$. Let $t_{jk}$ denote the travel time between the last node of $\sigma_1$ and the first node of $\sigma_2$. The concatenation rules are given by:
\begin{subequations}
\begin{align}
& T_D(\sigma^0) = s_k, \quad T_E(\sigma^0) = e_k, \quad T_L(\sigma^0) = l_k, \quad T_V(\sigma^0) = 0 \\
& \Delta_{t} = T_D(\sigma_1) + T_W(\sigma_1) + t_{jk} - T_V(\sigma_1) \\
& \Delta_{w} = \max\{T_E(\sigma_2) - \Delta_{t} - T_L(\sigma_1), 0\} \\
& \Delta_{v} = \max\{T_E(\sigma_1) + \Delta_{t} - T_L(\sigma_2), 0\} \\
& T_D(\sigma_1 \oplus \sigma_2) = T_D(\sigma_1) + t_{jk} + \Delta_{w} + T_D(\sigma_2) \label{eq:duration} \\
& T_E(\sigma_1 \oplus \sigma_2) = \max\{T_E(\sigma_1), T_E(\sigma_2) - \Delta_{t}\} - \Delta_{w} \\
& T_L(\sigma_1 \oplus \sigma_2) = \min\{T_L(\sigma_1), T_L(\sigma_2) - \Delta_{t}\} + \Delta_{v} \\
& T_V(\sigma_1 \oplus \sigma_2) = T_V(\sigma_1) + \Delta_{v} + T_V(\sigma_2)
\end{align}
\end{subequations}
By maintaining the attribute matrices, all concatenation operations can be computed in constant time, enabling efficient evaluation of attribute differences between the current solution and its neighbors within local search neighborhoods.
\section{Tensor-based GPU acceleration for local search operators}
\label{sec:method}
In this section, we present the Tensor-based GPU Acceleration (TGA) framework, which is designed to improve computational efficiency of commonly used local search operators in vehicle routing problems. The core lies in the design of the attribute-based solution tensor representation, and the tensor-based implementation of local search operators. By extracting and concatenating the relevant attribute tensors from the solution tensor representation, we can efficiently implement the sequence concatenation-based move evaluation with the help of GPU parallelism.

\subsection{Workflow of the TGA framework between CPU and GPU}
\label{sec:workflow}
The proposed TGA framework is designed with a low degree of coupling between the CPU and the GPU. The computationally intensive move evaluation of local search operators is entirely offloaded to the GPU. Meanwhile, the CPU handles only the updates based on the GPU's computation results. This design ensures that the TGA can be easily integrated into a wide range of local search-based algorithms and frameworks. 
\begin{figure}[!ht]
    \centering
    \includegraphics[width=0.9\textwidth]{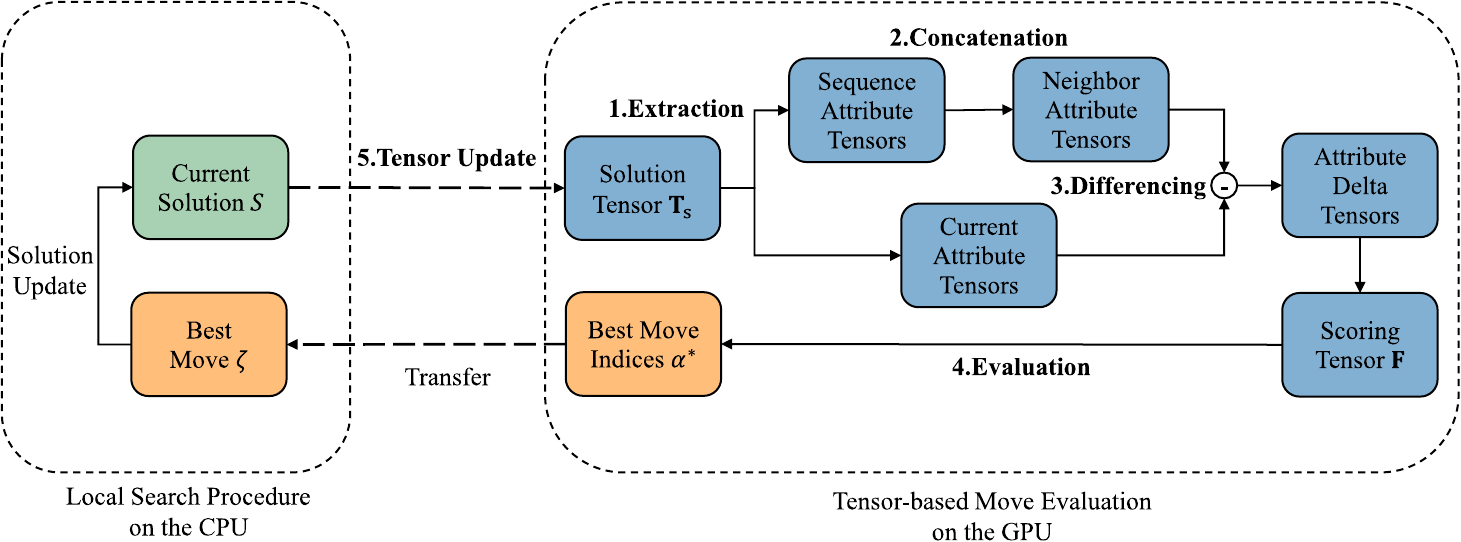}
    \caption{Workflow of the TGA framework.}
    \label{fig:workflow}
\end{figure}

The workflow of the TGA framework is illustrated in Figure~\ref{fig:workflow}. For a given solution $S$ to be improved by local search on the CPU, a solution tensor $\mathbf{T}_s$ is first initialized on the GPU using the attribute matrices derived from $S$. Tensor-based move evaluation is then performed on the GPU using $\mathbf{T}_s$. To fully exploit GPU parallelism, the best-improvement strategy is employed to select the best move among all candidate moves.

The tensor-based implementation of each operator follows a generic five-step process, the first four steps perform tensor-based move evaluation, and the final step performs the necessary updates. Specifically, these steps are: (1) \textbf{Extraction}: Extract relevant \emph{sequence attribute tensors} from the solution tensor $\mathbf{T}_s$ according to the specific operator. (2) \textbf{Concatenation}: Concatenate the extracted attribute tensors to construct the \emph{neighbor attribute tensors} which represent all neighboring solutions within the operator's neighborhood. (3) \textbf{Differencing}: Compute the \emph{attribute delta tensors} containing the differences between the current and modified routes. (4) \textbf{Evaluation}: Compute the \emph{scoring tensor} $\mathbf{F}$ based on the evaluation function and identify the best-improvement value and the corresponding indices $\alpha^*$ in $\mathbf{F}$ for the best move $\zeta$. (5) \textbf{Update}: Transfer the best move back to the CPU, update the solution $S$, and synchronize the updated solution tensor $\mathbf{T}_s$ along with other auxiliary tensors on the GPU for the next iteration.

\subsection{Attribute-based solution tensor representation}
\label{sec:solution_tensor}
\begin{figure}[!ht]
    \centering
    \includegraphics[width=0.8\textwidth]{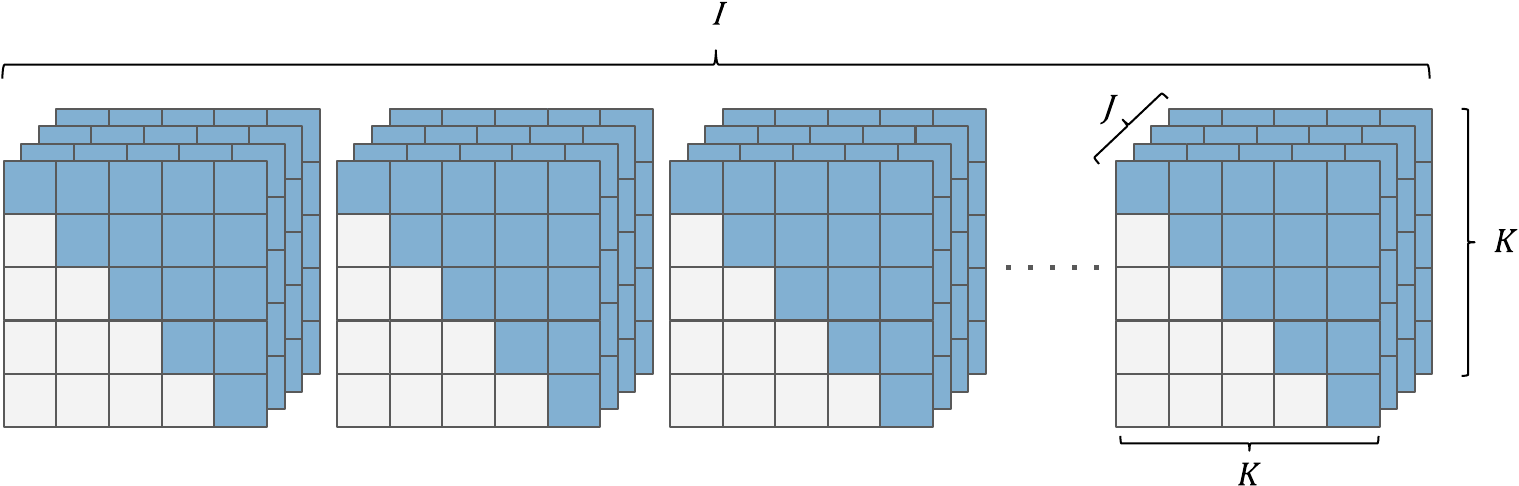}
    \caption{Illustration of the solution tensor representation.}
    \label{fig:tensor}
\end{figure}

As described in Section~\ref{sec:attribute_mat}, we use \emph{attribute matrices} to store attribute values, enabling constant-time move evaluation. Building on this, we aggregate these attribute matrices into a solution tensor $\mathbf{T}_s$ for GPU acceleration. Formally, the solution tensor $\mathbf{T}_s$ is a 4-dimensional tensor defined as
\begin{align}
\label{eq:sol_tensor}
\mathbf{T}_s \in \mathbb{R}^{I \times J \times K \times K} 
\end{align}
Here, $I$ denotes the number of maintained attributes, $J$ the maximum number of routes, and $K = \max_{\substack{1 \leq i \leq M}} \{|R_i|\}$ the maximum route length, i.e., the number of nodes in the longest route. The entries of $\mathbf{T}_s$ are indexed as
\begin{align}
\label{eq:sol_tensor}
(\mathbf{T}_s)_{ijkl}, \quad i \in \{1, \dots, I\}, j \in \{1, \dots, J\}, k,l \in \{1, \dots, K\}
\end{align}
where $(\mathbf{T}_s)_{ijkl}$ represents the value of the $i$-th attribute for the subsequence starting at position $k$ and ending at position $l$ in the $j$-th route. Figure~\ref{fig:tensor} illustrates the structure of the solution tensor representation.

To handle considered attributes introduced in Section~\ref{sec:calculation_concat}, we maintain the corresponding key attribute tensors: travel distance ($\mathbf{D}$), duration ($\mathbf{T}_D$), earliest and latest arrival times ($\mathbf{T}_E$, $\mathbf{T}_L$), time window violations ($\mathbf{T}_V$), incoming and outgoing loads ($\mathbf{L}_I$, $\mathbf{L}_O$), and maximum load ($\mathbf{L}_M$). Auxiliary tensors are also introduced to support efficient computation, including the first and last node indices of each sequence ($\mathbf{I}_S$, $\mathbf{I}_E$) and a boolean mask ($\mathbf{B}$) for valid entries. This attribute-based representation is flexible and can be extended for other VRP variants.

\subsection{Tensor-based implementation of local search operators}
\label{sec:tensor_operators}
Building on the solution tensor $\mathbf{T}_s$ defined in Equation~(\ref{eq:sol_tensor}), we develop the tensor-based implementation for four common local search operators in vehicle routing, \emph{Relocate}, \emph{Swap}, \emph{2-opt*}, and \emph{2-opt}, as introduced in Section~\ref{sec:sequence_concatenation}, taking advantage of GPU parallelism to speed up move evaluation. The following sections describe the five generic steps of the tensor-based implementation in detail.

\subsubsection{Extraction}
\label{sec:extraction}
Based on the sequence concatenation-based move evaluation introduced in Section~\ref{sec:move_eval}, \emph{Extraction} step extracts specific parts of the solution tensor $\mathbf{T}_s$, referred to as \emph{sequence attribute tensors}, using tensor operations such as indexing, slicing, and reshaping. These tensors contain all attribute values associated with the relevant route subsequences which are required to compute the attributes of the concatenated sequences in subsequent steps. This process leverages highly parallelizable tensor operations, such as slicing, indexing, and reshaping, making it well-suited for GPU acceleration. The \emph{Extraction} step is the critical step that distinguishes the operators and is implemented separately for \emph{intra-route} and \emph{inter-route} operators.
\paragraph{\textbf{Route-based extraction for intra-route operators}}
We propose a \emph{route-based extraction} for the intra-route operators (\emph{Intra-Relocate}, \emph{Intra-Swap} and \emph{2-opt}), which involve only a single route.

\begin{figure}[!htbp]
    \centering
    \includegraphics[width=0.8\textwidth]{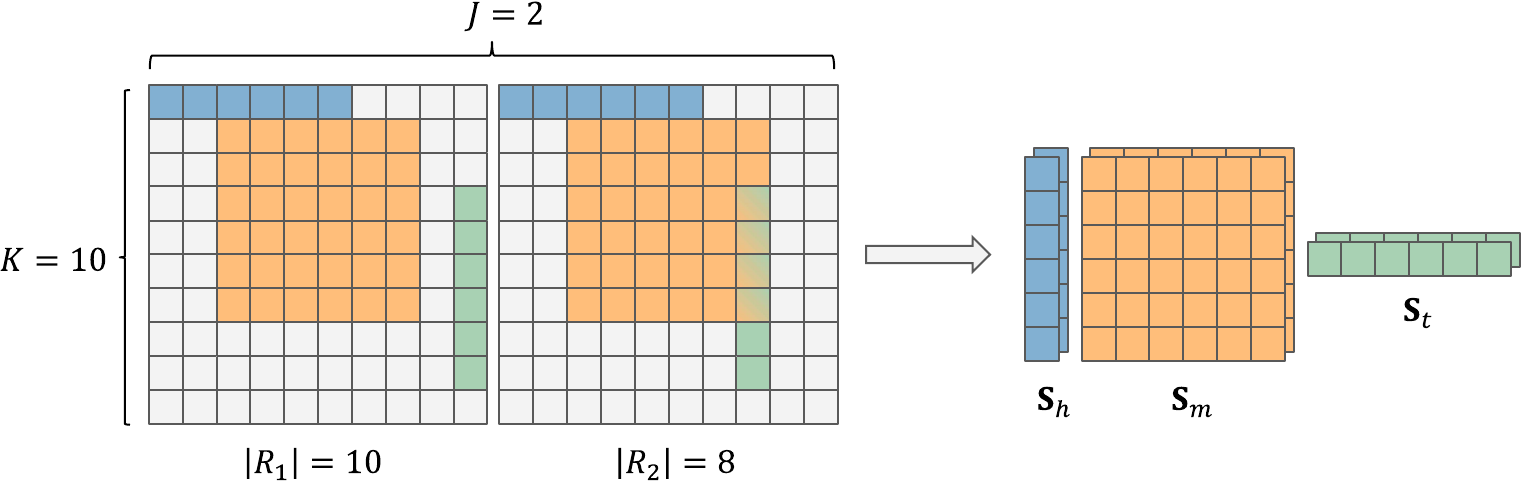}
    \caption{Example of the extraction operation for \emph{2-opt}. For clarity, we illustrate a single-attribute slice of the solution tensor and show a 3D tensor with dimensions $J \times K \times K = 2 \times 10 \times 10$, omitting the first dimension of $\mathbf{T}_s$.}
    \label{fig:ext_2_opt}
\end{figure}
\begin{figure}[!htbp]
    \centering
    \includegraphics[width=0.8\textwidth]{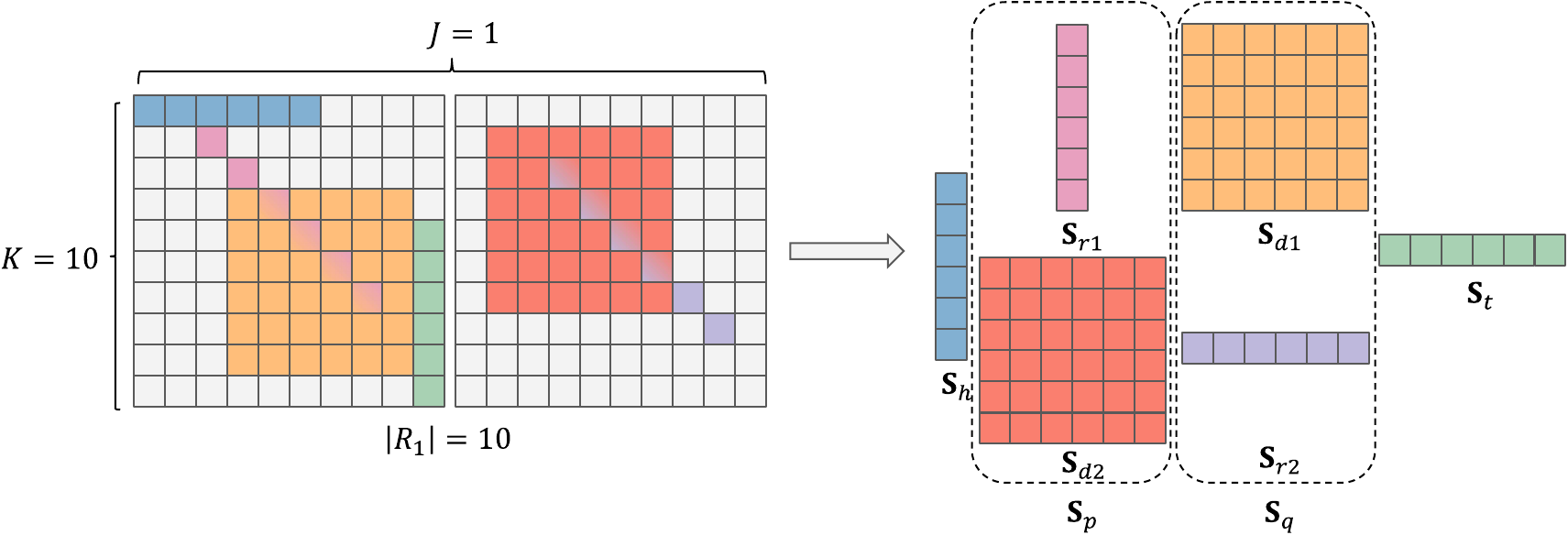}
    \caption{Example of the extraction operation for \emph{Intra-Relocate}. For clarity, we illustrate a single-attribute slice of the solution tensor and show a 3D tensor with dimensions $J \times K \times K = 1 \times 10 \times 10$, omitting the first dimension of $\mathbf{T}_s$ and setting $J=1$.}
    \label{fig:ext_intra_relocate}
\end{figure}
\begin{figure}[!htbp]
    \centering
    \includegraphics[width=0.8\textwidth]{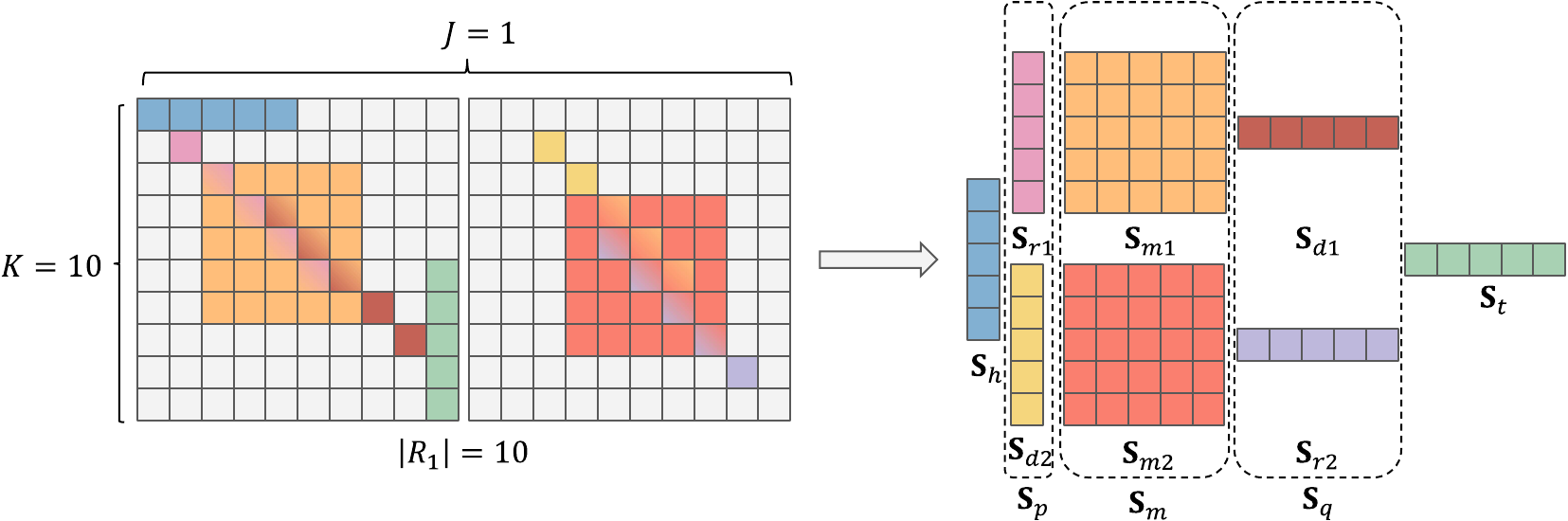}
    \caption{Example of the extraction operation for \emph{Intra-Swap} with $N_1=1, N_2=2$. For clarity, we illustrate a single-attribute slice of the solution tensor and show a 3D tensor with dimensions $J \times K \times K = 1 \times 10 \times 10$, omitting the first dimension of $\mathbf{T}_s$ and setting $J=1$.}
    \label{fig:ext_intra_swap}
\end{figure}
\begin{figure}[!htbp]
    \centering
    \includegraphics[width=0.8\textwidth]{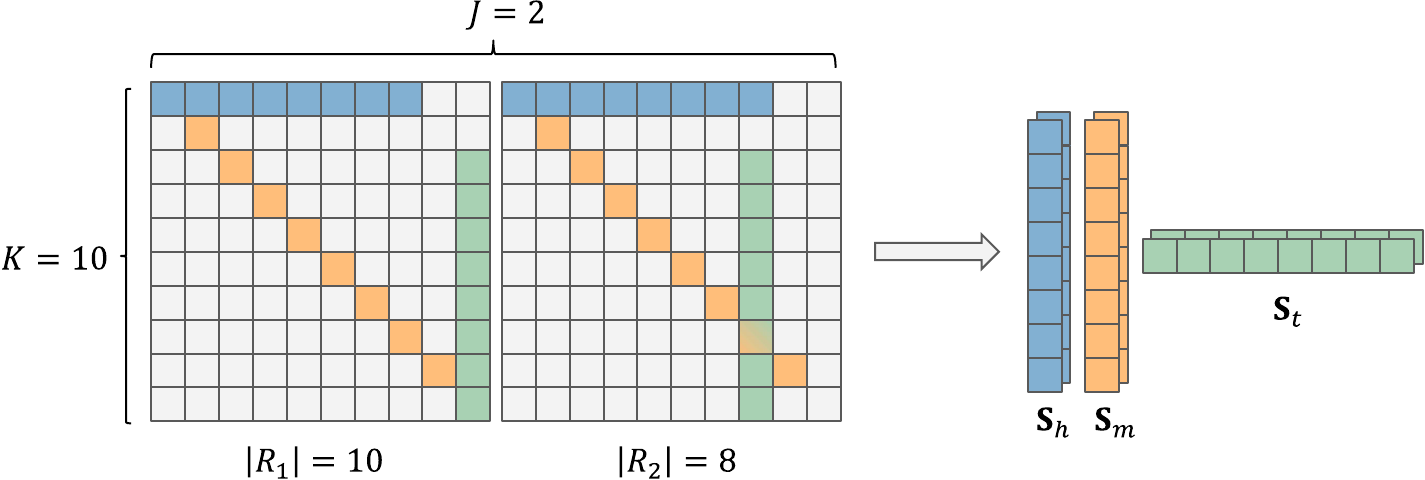}
    \caption{Example of the route-based \emph{3-Seq} operation with $N=1$ for inter-route operators. For clarity, we illustrate a single-attribute slice of the solution tensor and show a 3D tensor with dimensions $J \times K \times K = 2 \times 10 \times 10$. Note that the length of the routes affects the tail sequence~$\mathbf{S}_t$.}
    \label{fig:route_based_3seq}
\end{figure}
\begin{figure}[!htbp]
    \centering
    \includegraphics[width=\textwidth]{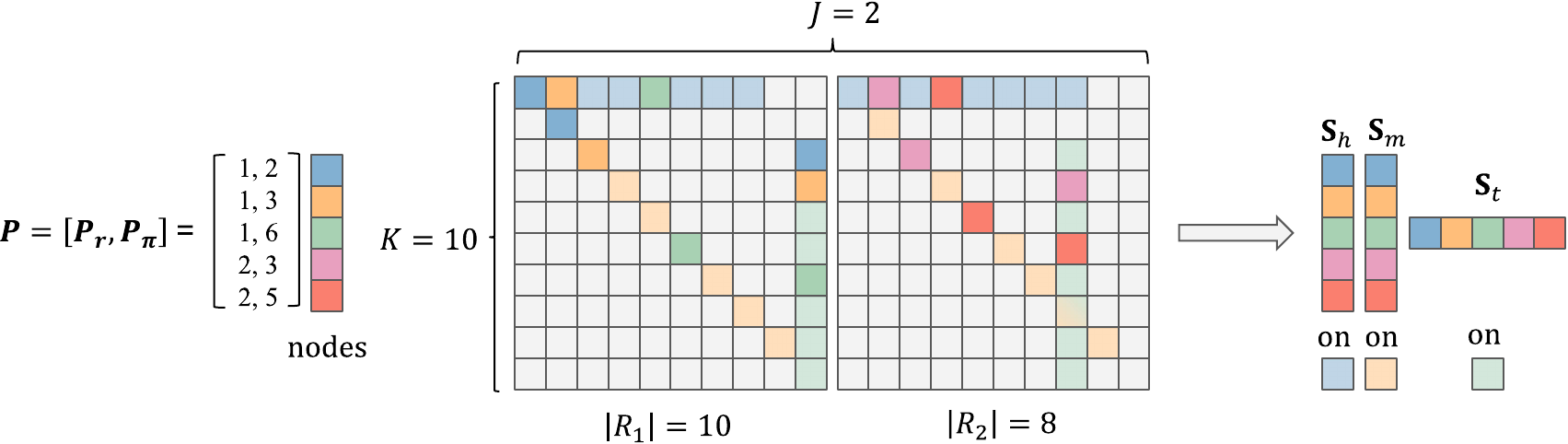}
    \caption{Example of the node-based \emph{3-Seq} operation. To illustrate the differences between node-based and route-based implementations, we reuse the same setting as in Figure~\ref{fig:route_based_3seq}. In this example, five nodes are distributed across two routes: the 2nd, 3rd, and 6th nodes on route $R_1$, and the 3rd and 5th nodes on route $R_2$. The figure shows how the head, middle, and tail sequences are extracted based on node positions, and highlights their relationship to those obtained through the route-based implementation.}
    \label{fig:node_based_3seq}
\end{figure}

As shown in Figure~\ref{fig:o-2opt}, the \emph{2-opt} operator divides the route at two cut points into three subsequences: the head sequence $S_h$, the middle sequence $S_m$, and the tail sequence $S_t$. The head sequence contains the nodes before the first cut, the middle sequence contains the nodes between the two cuts, and the tail sequence contains the nodes after the second cut. In the solution tensor $\mathbf{T}_s$, we extract the attribute values of all subsequences fully contained in each of these three parts into three \emph{sequence attribute tensors} $\mathbf{S}_h$, $\mathbf{S}_m$, and $\mathbf{S}_t$. Formally, for each attribute $i \in \{1, \dots, I\}$ and route $j \in \{1, \dots, J\}$, we define:
\begin{subequations}
\label{eq:2_opt}
\begin{align}
    &\mathbf{S}_h = (\mathbf{T}_s)_{ij1l}, \qquad l \in \{1, \dots, K-4\} \\
    &\mathbf{S}_m = (\mathbf{T}_s)_{ijkl}, \qquad k \in \{2, \dots, K-3\}, l \in \{3, \dots, K-2\} \\
    &\mathbf{S}_t = (\mathbf{T}_s)_{ijk|R_j|}, \qquad k \in \{4, \dots, K-1\} 
\end{align}
\end{subequations}
As illustrated in Figure~\ref{fig:ext_2_opt}, for each attribute $i$ and route $R_j$, the head sequence tensor $\mathbf{S}_h$ is always located in the first row of the attribute matrix, capturing the attribute values of relevant subsequences from the first node ($k=1$) up to the first cut point. The middle sequence tensor $\mathbf{S}_m$ stores attributes between two cut points, forming a square submatrix. The tail sequence tensor $\mathbf{S}_t$ stores attributes from second cut point to the last node, corresponding to the $R_j$-th column of the attribute matrix. The similar principles of the extraction process are illustrated for the other operators in the subsequent Figures~\ref{fig:ext_intra_relocate}-\ref{fig:node_based_3seq}. Since the \emph{2-opt} operator is only applied to VRP variants with symmetric settings, reversing the middle sequence can be implemented by swapping the first node index tensor $\mathbf{I_s}$ and the last node index tensor $\mathbf{I_e}$ of the attribute tensor $\mathbf{S}_m$. The reshaping operation is consistently applied to extracted \emph{sequence attribute tensors} across all operators, enabling the subsequent \emph{Concatenation} step via tensor broadcasting. 

As shown in Figure~\ref{fig:o-intra-relocate}, the \emph{Intra-Relocate} operator divides a route into four subsequences: the head sequence $S_h$, the relocated sequence $S_r$ of length $N$ ($N > 0$), the destination sequence $S_d$ preceding the insertion position, and the tail sequence $S_t$. The relative order of $S_r$ and $S_d$ depends on whether the relocated sequence is moved forward or backward along the route. To handle both cases efficiently on the GPU, we construct two combined sequence attribute tensors, $\mathbf{S}_p$ and $\mathbf{S}_q$. For each attribute $i \in \{1, \dots, I\}$ and route $j \in \{1, \dots, J\}$, the extraction proceeds as follows:
\begin{subequations}
\label{eq:ext_intra_relocate}
\begin{align}
&\mathbf{S}_h = (\mathbf{T}_s)_{ij1l}, \qquad l \in \{1, \dots, K-N-2\} \\ 
&\mathbf{S}_t = (\mathbf{T}_s)_{ijk|R_j|}, \qquad k \in \{N+3, \dots, K\} \\
&\mathbf{S}_p = \text{concat}(\mathbf{S}_{r1}, \mathbf{S}_{d2}), \qquad
\mathbf{S}_q = \text{concat}(\mathbf{S}_{d1}, \mathbf{S}_{r2}) 
\end{align}
\end{subequations}
The auxiliary tensors $\mathbf{S}_{r1}$, $\mathbf{S}_{r2}$, $\mathbf{S}_{d1}$, and $\mathbf{S}_{d2}$ capture the two possible relative positions of the relocated and destination sequences along the route. Specifically, $\mathbf{S}_{r1}$ and $\mathbf{S}_{d1}$ correspond to the case where the relocated sequence $S_r$ precedes the destination sequence $S_d$, while $\mathbf{S}_{r2}$ and $\mathbf{S}_{d2}$ correspond to the opposite case. These tensors are defined as follows. 
\begin{subequations}
\begin{align}
    &\mathbf{S}_{r1} = (\mathbf{T}_s)_{ijk(k+N-1)}, \qquad k \in \{2, \dots, K-N-1\} \\
    &\mathbf{S}_{d2} = (\mathbf{T}_s)_{ijkl}, \qquad k,l \in \{2, \dots, K-N-1\} \\
    &\mathbf{S}_{d1} = (\mathbf{T}_s)_{ijkl}, \qquad k,l \in \{N+2, \dots, K-1\} \\
    &\mathbf{S}_{r2} = (\mathbf{T}_s)_{ijk(k+N-1)}, \qquad k \in \{3, \dots, K-N\}
\end{align}
\end{subequations}
Similarly, as illustrated in Figure~\ref{fig:o-intra-swap}, the \emph{Intra-Swap} operator divides a single route into five subsequences: the head sequence $S_h$, the swapped sequence $S_r$ of length $N_1$ ($N_1 > 0$), the middle sequence $S_m$, the swapped sequence $S_d$ of length $N_2$ ($N_2 > 0$), and the tail sequence $S_t$. When the two swapped sequences have different lengths ($N_1 \neq N_2$), two cases must be considered, depending on whether $S_r$ or $S_d$ appears first. Thus, we also introduce auxiliary tensors to capture these cases. For each attribute $i \in \{1, \dots, I\}$ and route $j \in \{1, \dots, J\}$, the extraction process is defined as follows:
\begin{subequations}
\label{eq:ext_intra_swap}
\begin{align}
    &\mathbf{S}_h = (\mathbf{T}_s)_{ij1l}, \qquad l \in \{1, \dots, K-N_1-N_2-2\} \\
    &\mathbf{S}_t = (\mathbf{T}_s)_{ijk|R_j|}, \qquad k \in \{N_1+N_2+3, \dots, K\}, \\
    &\mathbf{S}_p = \operatorname{concat}(\mathbf{S}_{r1}, \mathbf{S}_{d2}), \qquad
    \mathbf{S}_m = \operatorname{concat}(\mathbf{S}_{m1}, \mathbf{S}_{m2}), \qquad
    \mathbf{S}_q = \operatorname{concat}(\mathbf{S}_{d1}, \mathbf{S}_{r2})
\end{align}
\end{subequations}
The auxiliary tensors $\mathbf{S}_{r1}$, $\mathbf{S}_{r2}$, $\mathbf{S}_{m1}$, $\mathbf{S}_{m2}$, $\mathbf{S}_{d1}$, and $\mathbf{S}_{d2}$ are defined as follows. Here, $\mathbf{S}_{r1}$, $\mathbf{S}_{m1}$ and $\mathbf{S}_{d1}$ correspond to the case where the swapped sequence $S_r$ precedes the swapped sequence $S_d$, while $\mathbf{S}_{r2}$, $\mathbf{S}_{m2}$ and $\mathbf{S}_{d2}$ correspond to the opposite case. 
\begin{subequations}
\begin{align}
    &\mathbf{S}_{r1} = (\mathbf{T}_s)_{ijk(k+N_1-1)}, \qquad k \in \{2, \dots, K-N_1-N_2-1\} \\
    &\mathbf{S}_{d2} = (\mathbf{T}_s)_{ijk(k+N_2-1)}, \qquad k \in \{2, \dots, K-N_1-N_2-1\} \\
    &\mathbf{S}_{m1} = (\mathbf{T}_s)_{ijkl}, \qquad k,l \in \{N_1+2, \dots, K-N_2-1\} \\
    &\mathbf{S}_{m2} = (\mathbf{T}_s)_{ijkl}, \qquad k,l \in \{N_2+2, \dots, K-N_1-1\} \\
    &\mathbf{S}_{d1} = (\mathbf{T}_s)_{ijk(k+N_2-1)}, \qquad k \in \{N_1+3, \dots, K-N_2\} \\
    &\mathbf{S}_{r2} = (\mathbf{T}_s)_{ijk(k+N_1-1)}, \qquad k \in \{N_2+3, \dots, K-N_1\}
\end{align}
\end{subequations}
\paragraph{\textbf{Route-based extraction for inter-route operators}}
Inter-route operators (\emph{Inter-Relocate}, \emph{Inter-Swap}, and \emph{2-opt*}) involve two distinct routes and share a common extraction scheme implemented via the \emph{3-Seq}($N$) operation, which splits the route into three subsequences: a head sequence $S_h$, a middle sequence $S_m$ of length $N$ to be relocated or swapped, and a tail sequence $S_t$. All three inter-route operators involve two \emph{3-Seq} operations, denoted as \emph{3-Seq}($N_1$) and \emph{3-Seq}($N_2$). Specifically, for \emph{Inter-Relocate}, $N_1 > 0$ and $N_2 = 0$; for \emph{Inter-Swap}, $N_1 > 0$ and $N_2 > 0$; and for \emph{2-opt*}, $N_1 = N_2 = 0$.

Building on the same principles used for intra-route operators, we extend the route-based extraction implementation to \emph{3-Seq} operation. For each attribute $i \in \{1, \dots, I\}$ and route $j \in \{1, \dots, J\}$, the route-based \emph{3-Seq($N$)} operation is defined as follows: 
\begin{subequations}
\begin{align}
\label{eq:inter_three_seq}
&\mathbf{S}_h = (\mathbf{T}_s)_{ij1l}, \qquad l \in \{1, \dots, K-N-1\}, N \ge 0 \\
&\mathbf{S}_m = (\mathbf{T}_s)_{ijk(k+N-1)}, \qquad k \in \{2, \dots, K-N\}, \text{ for } N > 0; \qquad \emptyset \text{ for } N = 0. \\
&\mathbf{S}_t = (\mathbf{T}_s)_{ijk|R_j|}, \qquad k \in \{N+2, \dots, K\}, N \ge 0
\end{align}
\end{subequations}
Let $L$ denote the feasible length of the \emph{sequence attribute tensors} after route-based extraction, with $L = K - N - 2$ for the route-based \emph{3-Seq} operation. Notably, this implementation generates high-dimensional intermediate and scoring tensors of size $O(J^2 L^2)$ that contain many invalid elements, resulting in substantial memory overhead and increased computational cost in subsequent computations. A detailed complexity analysis is provided in Section~\ref{sec:complexity}.
\paragraph{\textbf{Node-based extraction for inter-route operators}}
To address the inefficiencies of route-based extraction for inter-route operators, we propose a node-based implementation of the \emph{3-Seq} operation. This approach leverages explicit positional information to extract valid subsequences directly at the node level. Specifically, we construct \emph{positional tensors} $\mathbf{P} = \{\mathbf{P}_r, \mathbf{P}_\pi\}$ from the solution tensor $\mathbf{T}_s$, which encode the position of each valid node, including depots and customers. The \emph{route index tensor} $\mathbf{P}_r$ identifies the route to which each node belongs, while the \emph{position index tensor} $\mathbf{P}_\pi$ specifies the node's position within its route. Both $\mathbf{P}_r$ and $\mathbf{P}_\pi$ are one-dimensional tensors of size $Q$, where $Q = M + N_C$ denotes the total number of valid nodes across all routes.
\begin{subequations}
\label{eq:node_based1}
\begin{align}
&\mathbf{S}_h = (\mathbf{T}_s)_{i(p_r)k(p_\pi-1)} \\
&\mathbf{S}_m = (\mathbf{T}_s)_{i(p_r)(p_\pi)(p_\pi+N-1)} \\
&\mathbf{S}_t = (\mathbf{T}_s)_{i(p_r)(p_\pi+N)|R_{p_r}|}
\end{align}
\end{subequations}
Using this positional information, we extract the \emph{sequence attribute tensors} $\mathbf{S}_h$, $\mathbf{S}_m$, and $\mathbf{S}_t$ only for valid nodes, avoiding the invalid entries generated by the route-based extraction. Given positional indices $p_r \in \mathbf{P}_r$ and $p_\pi \in \mathbf{P}_\pi$, for each attribute $i \in \{1, \dots, I\}$, the node-based extraction distinguishes two cases. If $N > 0$, the middle sequence exists (Equation~(\ref{eq:node_based1})); if $N = 0$, the middle sequence is empty (Equation~(\ref{eq:node_based2})).
\begin{subequations}
\label{eq:node_based2}
\begin{align}
&\mathbf{S}_h = (\mathbf{T}_s)_{i(p_r)k(p_\pi)} \\
&\mathbf{S}_m = \emptyset \\
&\mathbf{S}_t = (\mathbf{T}_s)_{i(p_r)(p_\pi+1)|R_{p_r}|}
\end{align}
\end{subequations}
Unlike the route-based extraction, which is based on the route structure and produces high-dimensional tensors with many invalid elements, the node-based extraction focuses on valid nodes using their positional information, resulting in greater computational efficiency. As a result, the intermediate and scoring tensors are reduced to size $O(Q^2)$, significantly lowering memory consumption and computational cost and enabling a more efficient GPU implementation.
\paragraph{\textbf{Edge-based extraction for inter-route operators}}
\label{sec:edge_based_extraction}
Although the node-based extraction significantly reduces memory usage and computational complexity compared to the route-based method, it still evaluates the entire neighborhood and therefore cannot directly leverage neighborhood reduction strategies \cite{vidal2013hybrid} commonly used in CPU-based local search, which restrict neighborhood evaluation to promising moves based on specific criteria to improve search efficiency. To overcome this limitation, we propose an edge-based extraction as one possible implementation that shifts the focus from nodes to edges connecting pairs of nodes. 

Building on the \emph{positional tensors} $\mathbf{P}=\{\mathbf{P}_r, \mathbf{P}_\pi\}$ introduced in the node-based extraction, this edge-based method enables selective evaluation of promising moves while avoiding unnecessary computations. Specifically, we introduce a binary mask tensor $\mathbf{M} \in \{0,1\}^{|\mathcal{V}| \times |\mathcal{V}|}$ to encode the outcome of neighborhood reduction. An entry $\mathbf{M}_{ij}=1$ indicates that the edge $(v_i, v_j)$ is valid under a given pruning criterion, and $\mathbf{M}_{ij}=0$ otherwise. This mask restricts move evaluation to a reduced set of node pairs. Using the mask tensor $\mathbf{M}$ together with the positional tensor $\mathbf{P}$, we retain only valid node pairs $\{(v_i, v_j) \mid \mathbf{M}_{ij}=1\}$ and construct the corresponding \emph{edge index tensors} $\mathbf{E} = \{\mathbf{E}_i, \mathbf{E}_j\}$. For each valid edge $(v_i,v_j)$, the tensors $\mathbf{E}_i$ and $\mathbf{E}_j$ store the positional information of the source and target nodes, respectively, obtained by indexing $\mathbf{P}$ as
\begin{align}
\mathbf{E}_i = \mathbf{P}_i = \{\mathbf{P}_{r_i}, \mathbf{P}_{\pi_i}\},
\quad
\mathbf{E}_j = \mathbf{P}_j = \{\mathbf{P}_{r_j}, \mathbf{P}_{\pi_j}\},
\quad \forall (i,j)\ \text{such that } \mathbf{M}_{ij}=1.
\end{align}
Based on these \emph{edge index tensors}, the edge-based implementation of \emph{3-Seq} operation extracts the \emph{sequence attribute tensors} $\mathbf{S}_h$, $\mathbf{S}_m$, and $\mathbf{S}_t$ following the same procedure as in the node-based extraction. Let $E$ denote the number of selected edges (i.e., the number of nonzero entries in $\mathbf{M}$). The resulting intermediate and scoring tensors have size $O(E)$, yielding substantial reductions in memory consumption and computational cost compared to the node-based method.

In summary, we consider three extraction strategies for inter-route operators: route-based, node-based, and edge-based. As node-based extraction is more efficient than route-based (see the detailed analysis in Section~\ref{sec:complexity}), we define two standard TGA implementations: the node-implemented TGA (NTGA), which applies node-based extraction to inter-route operators with full neighborhood evaluation, and the edge-implemented TGA (ETGA), which uses edge-based extraction to enable neighborhood reduction. Both implementations use route-based extraction for intra-route operators.
\subsubsection{Concatenation}
\label{sec:concatenation}
After extracting the \emph{sequence attribute tensors}, the extracted tensors are sliced along the first dimension to obtain the individual attribute tensors. The concatenation operation is then applied to these attribute tensors to construct the \emph{neighbor attribute tensors}, which contain the attributes of the concatenated sequences and represent the neighboring solutions generated by the candidate moves.
\begin{subequations}
\begin{align}
\label{eq:concat_distance}
&\mathbf{D}(\mathbf{S}_1 \oplus \mathbf{S}_2) = \mathbf{D}(\mathbf{S}_1) + \mathbf{C}_{ij} + \mathbf{D}(\mathbf{S}_2) \\
\label{eq:concat_duration}
&\mathbf{T}_D(\mathbf{S}_1 \oplus \mathbf{S}_2) = \mathbf{T}_D(\mathbf{S}_1) + \mathbf{T}_{ij} + \mathbf{\Delta}_w + \mathbf{T}_D(\mathbf{S}_2) \\
&i \in \mathbf{I}_e(\mathbf{S}_1), \quad j \in \mathbf{I}_s(\mathbf{S}_2)
\end{align}
\end{subequations}
Let $\mathbf{S}_1$ and $\mathbf{S}_2$ denote the \emph{sequence attribute tensors} of sequences $\sigma_1$ and $\sigma_2$, respectively. Most attributes of the concatenated sequence can be computed directly using the tensorized formulas in Section~\ref{sec:calculation_concat}, which only involve basic arithmetic operations and are therefore well-suited for GPU parallelization. However, the computation of travel distance and duration requires accounting for the additional costs of connecting the two sequences. Specifically, these costs can be derived from the travel distance tensor $\mathbf{C}$ (corresponding to the travel distance matrix $\mathcal{C}$), the travel time tensor $\mathbf{T}$ (corresponding to the travel time matrix $\mathcal{T}$), the last node index tensor of $\mathbf{S}_1$ ($\mathbf{I}_e(\mathbf{S}_1)$), and the first node index tensor of $\mathbf{S}_2$ ($\mathbf{I}_s(\mathbf{S}_2)$). The resulting concatenation formulas for travel distance $\mathbf{D}$ and duration $\mathbf{T}_D$ are given in Equations~(\ref{eq:concat_distance}) and~(\ref{eq:concat_duration}), respectively, which correspond to Equations~(\ref{eq:dist}) and~(\ref{eq:duration}) in Section~\ref{sec:calculation_concat}.

\subsubsection{Differencing}
\label{sec:differencing}
To evaluate the effect of a candidate move, we compute the changes in key attributes between the current and modified routes. The attributes of the current routes are stored in the solution tensor $\mathbf{T}_s$, referred to as the \emph{current attribute tensors}, while the attributes of the modified routes are obtained through the \emph{Concatenation} step, referred to as \emph{neighbor attribute tensors}. The differences are then computed element-wise to produce the \emph{attribute delta tensors}, which include the changes in travel distance ($\Delta \mathbf{D}$), time window violations ($\Delta \mathbf{T}_V$), and load violations ($\Delta \mathbf{L}_V$). These attribute delta tensors form the input for the \emph{Evaluation} step, enabling the tensorized computation of the \emph{scoring tensor} $\mathbf{F}$ for all candidate moves.
\subsubsection{Evaluation}
\label{sec:evaluation}
Building on the \emph{attribute delta tensors} computed in the \emph{Differencing} step, the \emph{scoring tensor} $\mathbf{F}$ is calculated using the defined evaluation function $\mathcal{F}$ (Equation~(\ref{eq:scoring_tensor})). Invalid entries are filtered via the boolean mask tensor $\mathbf{B}$ by assigning a large penalty (Equation~(\ref{eq:mask})).
\begin{subequations}
   \begin{align} 
    \label{eq:scoring_tensor}
    \mathbf{F} &= \mathcal{F}(\Delta \mathbf{D}, \Delta \mathbf{T}_V, \Delta \mathbf{L}_V) \\
    \label{eq:mask}
    \mathbf{F} &\leftarrow \mathbf{F} + (\neg \mathbf{B}) \cdot \infty \\
    \label{eq:min_obj}
    F_{\text{min}} &= \mathbf{F}_{\boldsymbol{\alpha}^*} \quad \text{where} \quad \boldsymbol{\alpha}^* = \arg\min_{\boldsymbol{\alpha}} \mathbf{F}_{\boldsymbol{\alpha}}
\end{align} 
\end{subequations}
Then, an \emph{argmin} operation is applied to the \emph{scoring tensor} $\mathbf{F}$ to identify the best move with the minimal score value $F_{\text{min}}$ and its corresponding indices $\boldsymbol{\alpha}^*$ (Equation~(\ref{eq:min_obj})). This operation is realized via a parallel \emph{reduction} on the GPU. The indices $\boldsymbol{\alpha}^*$ encode the information necessary to identify the best move, including the involved routes and the starting positions of the sequences to be relocated or swapped. Finally, this move information is transmitted to the CPU and used to update the solution $S$.
\subsubsection{Tensor update}
\label{sec:tensor_update}
Once the best move is performed on the CPU, the solution tensor representation $\mathbf{T}_s$ needs to be updated based on the resulting solution. Since the considered operators only impact one or two routes, only the affected parts of the tensor representation $\mathbf{T}_s$ require updating, avoiding the need to reconstruct the entire tensor from scratch. Specifically, the update process consists of two main steps: (1) Resize the solution tensor $\mathbf{T}_s$ if the maximum number of routes $J$ or the maximum route length $K$ change. (2) Update the attribute values of the affected routes on the tensor representation $\mathbf{T}_s$ on GPU. Additionally, the \emph{positional tensors} $\mathbf{P} = \{\mathbf{P}_r, \mathbf{P}_\pi\}$ and the \emph{edge index tensors} $\mathbf{E} = \{\mathbf{E}_i, \mathbf{E}_j\}$ also need to be updated accordingly if the node-based or edge-based extraction methods are used for inter-route operators.

\subsection{Complexity analysis}
\label{sec:complexity}
We analyze the space and time complexity of the proposed tensor-based GPU acceleration (TGA) framework in terms of the number of tensor elements processed. We assume that up to $p$ GPU threads can be scheduled in parallel. Let $I$ be the number of maintained attributes, $J$ the maximum number of routes, and $K$ the maximum route length. Let $L$ denote feasible length of the \emph{sequence attribute tensors} in the route-based extraction. For inter-route operators, let $Q = M + N_C$ be the total number of valid nodes in the node-based extraction, and $E$ the number of valid node pairs after neighborhood reduction in the edge-based extraction.
\paragraph{\textbf{Tensor initialization and update}}
The solution tensor $\mathbf{T}_s \in \mathbb{R}^{I \times J \times K \times K}$ requires $O(IJK^2)$ space and can be constructed on the GPU in $O(IJK^2/p)$ time. Since each accepted move modifies at most two routes, updating $\mathbf{T}_s$ costs $O(IK^2)$ space and $O(IK^2/p)$ time.

When node-based or edge-based extraction is enabled for inter-route operators, the \emph{positional tensors} $\mathbf{P}=\{\mathbf{P}_r,\mathbf{P}_\pi\}$ and the \emph{edge index tensors} $\mathbf{E}=\{\mathbf{E}_i,\mathbf{E}_j\}$ must be updated. Updating the \emph{positional tensors} $\mathbf{P}$ has the complexity of $O(Q)$, whereas updating the \emph{edge index tensors} $\mathbf{E}$ requires generating valid node pairs based on the \emph{positional tensors} $\mathbf{P}$ and the mask tensor $\mathbf{M}$, and has a complexity of $O(Q^2)$, potentially becoming the bottleneck in the edge-based implementation.

Additionally, in practice, tensor initialization and updates involve data transfers between the CPU and GPU for building or updating the solution tensor $\mathbf{T}_s$. As a result, when tensor-based evaluation is relatively fast or the neighborhood size is small, data transfer overheads may dominate the total runtime despite the favorable complexity.

\paragraph{\textbf{Tensor-based move evaluation}}
Tensor-based move evaluation consists of four steps: \emph{Extraction}, \emph{Concatenation}, \emph{Differencing}, and \emph{Evaluation}. The \emph{Extraction} step performs slicing, indexing, and reshaping operations on the solution tensor $\mathbf{T}_s$ to obtain the \emph{sequence attribute tensors}. The \emph{Concatenation} step constructs \emph{neighbor attribute tensors} for each attribute independently, while the \emph{Differencing} step computes \emph{attribute delta tensors} for required attributes according to the evaluation function. Both of them involve element-wise arithmetic operations and broadcasting. These tensor operations involved in the first three steps can be highly parallelized on GPU. The Evaluation step computes the \emph{scoring tensor} $\mathbf{F}$ from the \emph{attribute delta tensors} and identifies the best move via a \emph{reduction} operation. This step has a time complexity of $O(n/p + \log p)$, where $n$ denotes the number of elements in $\mathbf{F}$.

Based on the above analysis, Table~\ref{tab:complexity_steps} presents the dominant overall complexity and  summarizes the per-step space and time complexity for route-based intra-route operators and for three extraction strategies of inter-route operators: route-based, node-based, and edge-based.
\begin{table}[!ht]
\centering
\caption{Per-step space and time complexity of tensor-based move evaluation for intra-route operators (route-based), and inter-route operators (route-based, node-based, and edge-based).}
\label{tab:complexity_steps}
\scriptsize
\setlength{\tabcolsep}{3.2pt}
\resizebox{\textwidth}{!}{%
\begin{tabular}{lcccccccc}
\hline
\multirow{2}{*}{Step}
	& \multicolumn{2}{c}{Intra-route (route-based)}
	& \multicolumn{2}{c}{Inter-route (route-based)}
	& \multicolumn{2}{c}{Inter-route (node-based)}
	& \multicolumn{2}{c}{Inter-route (edge-based)} \\
\cmidrule(l){2-3} \cmidrule(l){4-5} \cmidrule(l){6-7} \cmidrule(l){8-9}
 & Space & Time & Space & Time & Space & Time & Space & Time \\
\hline
Extraction & $O(IJL^2)$ & $O\!\left(\frac{IJL^2}{p}\right)$ & $O(IJL)$ & $O\!\left(\frac{IJL}{p}\right)$ & $O(IQ)$ & $O\!\left(\frac{IQ}{p}\right)$ & $O(IE)$ & $O\!\left(\frac{IE}{p}\right)$ \\
Concatenation & $O(IJL^2)$ & $O\!\left(I\frac{JL^2}{p}\right)$ & $O(IJ^2L^2)$ & $O\!\left(I\frac{J^2L^2}{p}\right)$ & $O(IQ^2)$ & $O\!\left(I\frac{Q^2}{p}\right)$ & $O(IE)$ & $O\!\left(I\frac{E}{p}\right)$ \\
Differencing & $O(JL^2)$ & $O\!\left(\frac{JL^2}{p}\right)$ & $O(J^2L^2)$ & $O\!\left(\frac{J^2L^2}{p}\right)$ & $O(Q^2)$ & $O\!\left(\frac{Q^2}{p}\right)$ & $O(E)$ & $O\!\left(\frac{E}{p}\right)$ \\
Evaluation & $O(JL^2)$ & $O\!\left(\frac{JL^2}{p}+\log p\right)$ & $O(J^2L^2)$ & $O\!\left(\frac{J^2L^2}{p}+\log p\right)$ & $O(Q^2)$ & $O\!\left(\frac{Q^2}{p}+\log p\right)$ & $O(E)$ & $O\!\left(\frac{E}{p}+\log p\right)$ \\
\hline
Overall & $O(IJL^2)$ & $O\!\left(I\frac{JL^2}{p}+\log p\right)$ & $O(IJ^2L^2)$ & $O\!\left(I\frac{J^2L^2}{p}+\log p\right)$ & $O(IQ^2)$ & $O\!\left(I\frac{Q^2}{p}+\log p\right)$ & $O(IE)$ & $O\!\left(I\frac{E}{p}+\log p\right)$ \\
\hline
\end{tabular}
}
\end{table}
The table shows that the overall computational cost is largely determined by the size of the \emph{scoring tensor} $\mathbf{F}$. When $\mathbf{F}$ is large, the \emph{Evaluation} step dominates the computational cost; otherwise, the \emph{Extraction} and \emph{Concatenation} steps, particularly the latter, become the main contributors. For inter-route operators, the sizes satisfy $J^2L^2 > Q^2 > E$, due to the presence of the large number of invalid entries in the route-based implementation and the neighborhood reduction in the edge-based implementation. Consequently, the route-based implementation exhibits the highest space and time complexity. The node-based implementation significantly reduces complexity by restricting evaluation to valid nodes, while the edge-based implementation further improves computational efficiency by limiting evaluation to $E$ valid node pairs, which is typically much smaller than $Q^2$.

However, the edge-based implementation introduces additional overhead associated with constructing and maintaining the \emph{edge index tensors} $\mathbf{E}$. Moreover, when neighborhood reduction is less effective, this approach may compromise solution quality. These trade-offs are examined in detail in the next section.

\section{Computational results}
\label{sec:computational_results}
This section presents computational experiments to evaluate the performance of the proposed tensor-based GPU acceleration (TGA) framework, focusing on two implementations: the node-implemented TGA (NTGA) and the edge-implemented TGA (ETGA), applied to common local search operators in vehicle routing problems.

It is important to note that the goal of this study is not to set new records for the benchmark instances. Instead, the goal is to explore the potential of the proposed TGA framework for solving vehicle routing problems and to provide a foundation for future research in this direction. Accordingly, the following sections focus on designing fair and consistent experiments to assess the effectiveness of the TGA framework, particularly in terms of computational efficiency, compared to traditional CPU-based implementations.

\subsection{Algorithms for experimentation}
\label{sec:algorithm}
This section describes the algorithms used to evaluate the proposed TGA framework. TGA is highly extensible and can be integrated into various local search-based algorithms. To demonstrate its effectiveness, we incorporated TGA into the MA-FIRD algorithm \cite{lei2024memetic}, which has shown its effectiveness in solving the VRPSPDTW. Since the CVRP and VRPTW are special cases of the VRPSPDTW, this integration allows the algorithm to address these variants as well, even though MA-FIRD was not originally designed for them. MA-FIRD employs a population-based evolutionary framework, feasible and infeasible search, the best-improvement local search strategy, and sequence concatenation-based move evaluation. 

To evaluate TGA, we first adapted the original MA-FIRD algorithm of \cite{lei2024memetic} as the baseline, denoted MA-N. Its TGA-based counterpart, referred to as MA-NTGA, integrates the proposed node-implemented TGA (NTGA) into MA-N. Both MA-N and MA-NTGA perform full neighborhood evaluation without applying neighborhood reduction of Section \ref{sec:edge_based_extraction}. To further assess the impact of neighborhood reduction in conjunction with TGA, we adopted two additional algorithm variants (MA-E and MA-ETGA) using the neighborhood reduction mechanism proposed in \cite{vidal2013hybrid}, with the granularity threshold $\theta$ controlling the number of nearest neighbors per node. Specifically, MA-E extends the CPU-based MA-N by incorporating neighborhood reduction into its local search operators, while MA-ETGA is its TGA-based counterpart, which integrates the edge-implemented TGA (ETGA) to support the same neighborhood reduction mechanism.
\begin{table}[!htbp]
\centering
\caption{Characteristics of the four algorithm variants.}
\label{tab:algo_variants}
\begin{tabular}{lcc}
\hline
Algorithm & Move Evaluation & Neighborhood Reduction \\ \hline
MA-N     & CPU-based & No  \\
MA-E     & CPU-based  & Yes \\
MA-NTGA  & TGA-based & No  \\
MA-ETGA  & TGA-based & Yes \\
\hline
\end{tabular}
\end{table}

As summarized in Table~\ref{tab:algo_variants}, these four algorithms (MA-N, MA-E, MA-NTGA, and MA-ETGA) share the same memetic algorithm framework, best-improvement local search strategy, and set of operators: \emph{Relocate}, \emph{Swap}, \emph{2-opt*} for the VRPTW and VRPSPDTW, plus \emph{2-opt} for the CVRP. The only differences among them lie solely in their move evaluation implementations and the use of neighborhood reduction. A detailed description of these algorithms is provided in Section~A2 of the supplementary material.

\subsection{Benchmark instances}
\label{sec:benchmark_instances}
We assess the proposed TGA on three VRP variants: the capacitated VRP (CVRP), VRP with time windows (VRPTW), and the VRP with simultaneous pickup and delivery and time windows (VRPSPDTW), using widely recognized benchmark instances from the literature. Specifically, experiments were conducted on a representative subset of 210 benchmark instances across three VRP variants: 100 \emph{X} benchmark instances \cite{uchoa2017new} for CVRP, with 100 to 1000 customers; 20 \emph{JD} benchmark instances \cite{liu2021memetic} for VRPSPDTW, based on real-world JD Logistics data with 200, 400, 600, 800, and 1000 customers; and 90 selected \emph{GH} instances for VRPTW, covering a wide range of sizes and characteristics. The \emph{GH} benchmark \cite{gehring1999parallel} originally comprises 300 instances with 200 to 1000 customers, divided into three types: \emph{C} (clustered), \emph{R} (random), and \emph{RC} (mixed), each including instances with both narrow and wide time windows and small and large vehicle capacities.

\subsection{Experimental protocol}
\label{sec:exp_protocal}
All four algorithm variants, MA-N, MA-E, MA-NTGA, and MA-ETGA, were implemented in C++, with the tensor-based GPU acceleration leveraging the PyTorch C++ API. The source code will be made publicly available upon publication of this paper. Experiments were performed on an Intel Xeon Gold 6248 2.50 GHz processor and an NVIDIA V100 GPU with 32 GB memory. As discussed in Section~\ref{sec:gpu_architecture}, the V100 is no longer state-of-the-art but remains a mature and widely used GPU platform for algorithm evaluation. Importantly, since TGA is deterministic and hardware-independent, the GPU choice affects only runtime performance, and newer GPUs are expected to further improve efficiency without changing solution quality.

Our experimental design aims to provide a fair and consistent evaluation of TGA, with a particular focus on computational gains over traditional CPU-based implementations. To ensure controlled comparisons, the algorithmic parameters were adopted from MA-FIRD \cite{lei2024memetic} and kept consistent across all variants, avoiding biases due to differing configurations. Key parameters, including the population size $\mu$ and the granularity threshold $\theta$ (introduced in MA-E and MA-ETGA), were tuned using the automatic parameter configuration tool Irace \cite{lopez2016irace} on a selected subset of instances from each benchmark. Table~\ref{tab:params} reports the candidate ranges and selected values for each benchmark set.
\begin{table}[!htbp]
\centering
\caption{Parameter tuning results.}
\label{tab:params}
\begin{tabular}{llllll}
\hline
Parameter & Description & Candidate values & \emph{X} & \emph{GH} & \emph{JD} \\ \hline
$\mu$     & Population size        & \{10, 20, 30, 40, 50\}     & 40 & 10  & 10 \\
$\theta$  & Granularity threshold  & \{20, 50, 80, 100, 120\}  & 20 & 100 & 80 \\ \hline
\end{tabular}
\end{table}

To ensure fair comparisons with state-of-the-art methods, we adopt the same termination criteria as in the corresponding studies for each benchmark set. For the \emph{X} (CVRP) instances, the time limit is proportional to the number of customers, set to $N_C \times 240 / 100$ seconds \cite{vidal2012hybrid,vidal2020hybrid}. For the \emph{GH} (VRPTW) instances, the limits are 1800, 3600, and 7200 seconds for instances with $\le 200$, $\le 800$, and $>800$ customers, respectively \cite{kool2022hybrid}. For the \emph{JD} (VRPSPDTW) instances, following \cite{lei2024memetic}, termination occurs when any of the following criteria is met: a maximum of 5000 generations, 500 consecutive generations without improvement, or a time limit of 7200 seconds. Each instance was solved 10 times by each algorithm variant.

\subsection{Computational results and comparisons}
\label{sec:exp_results}
\begin{figure}[!htbp]
\centering
\includegraphics[width=0.6\textwidth]{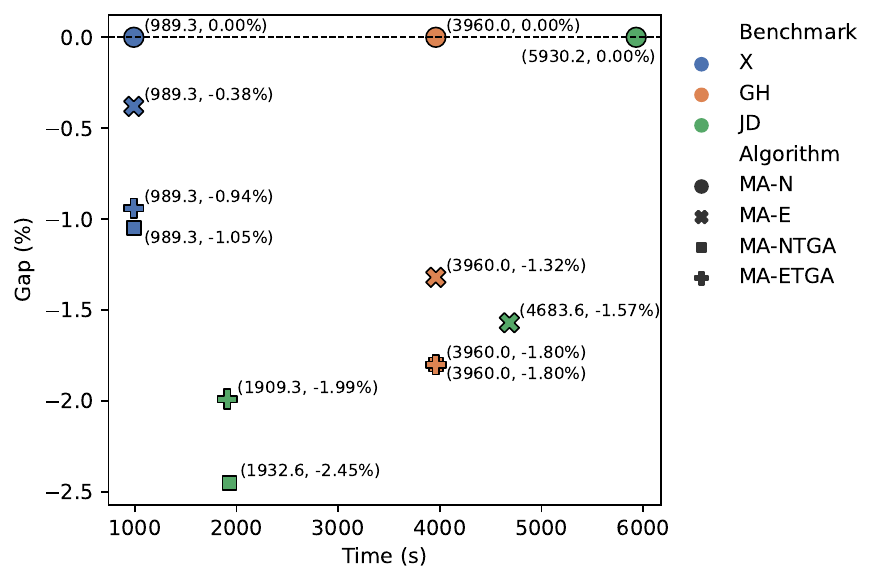}
\caption{Comparative performance of MA-N, MA-E, MA-NTGA, and MA-ETGA on three benchmark sets. The x-axis shows average computation time (s), and the y-axis shows average gap (\%) relative to the baseline MA-N.}
\label{fig:summ_plot}
\end{figure}

Figure~\ref{fig:summ_plot} compares the average computation time and average gap of MA-E, MA-NTGA, and MA-ETGA relative to the baseline MA-N across 210 instances of three benchmark sets: \emph{X} (CVRP), \emph{GH} (VRPTW), and \emph{JD} (VRPSPDTW). The average gap is defined as the relative difference in objective value with respect to MA-N, with detailed numerical results reported in Tables~A3-A7 in the supplementary material. 
Overall, all three variants outperform MA-N, while the TGA-based algorithms further improve upon both MA-N and MA-E across all benchmarks. For the \emph{X} and \emph{GH} instances, where termination is solely time-based, performance differences are clearly reflected in the gap values. On \emph{X}, MA-NTGA and MA-ETGA achieve average gaps of -1.05\% and -0.94\%, respectively, compared to -0.38\% for MA-E. On \emph{GH}, both MA-NTGA and MA-ETGA attain average gaps of -1.80\%, outperforming MA-E's -1.32\%. For the \emph{JD} instances, which additionally use generation-based and stagnation-based stopping criteria, MA-NTGA and MA-ETGA achieve larger improvements (-2.45\% and -1.99\%) than MA-E (-1.57\%), while requiring shorter average computation times. These results demonstrate both the computational efficiency and enhanced search capability of the TGA-based implementations.

MA-NTGA and MA-ETGA exhibit comparable performance, with MA-NTGA showing a slight advantage. This suggests that although the edge-implemented TGA (ETGA) effectively accelerates move evaluation by reducing the neighborhood size, it does not automatically translate into superior solution quality. The effectiveness of neighborhood reduction strongly depends on its design, inappropriate pruning may eliminate promising high-quality moves. Moreover, the edge-based implementation introduces extra overhead, affecting overall efficiency, which is analyzed in detail in the following section.

\subsection{Analysis of overall speedup ratio}
\label{sec:analysis_speedup}
\begin{figure}[!htbp]
    \centering
    \begin{subfigure}{0.46\textwidth}
        \centering
        \includegraphics[width=\textwidth]{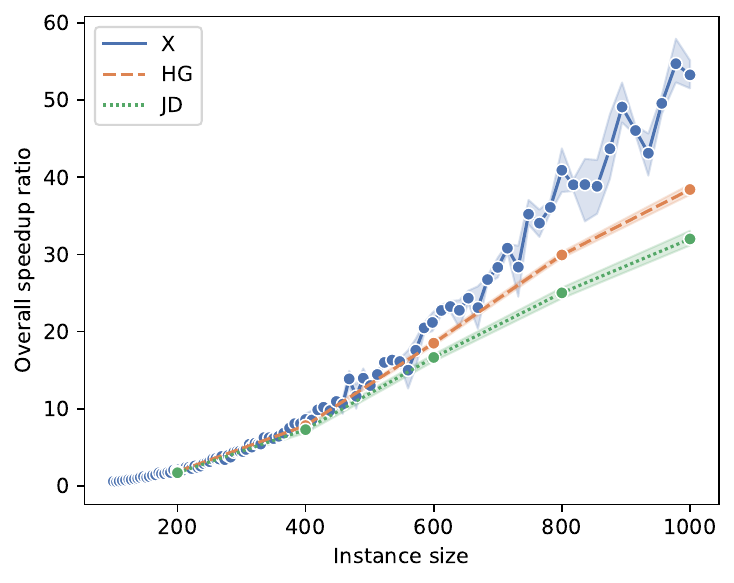}
        \caption{\small Overall speedup of MA-NTGA relative to MA-N.}
        \label{fig:speedup_ntga}
    \end{subfigure}
    \hfill
    \begin{subfigure}{0.46\textwidth}
        \centering
        \includegraphics[width=\textwidth]{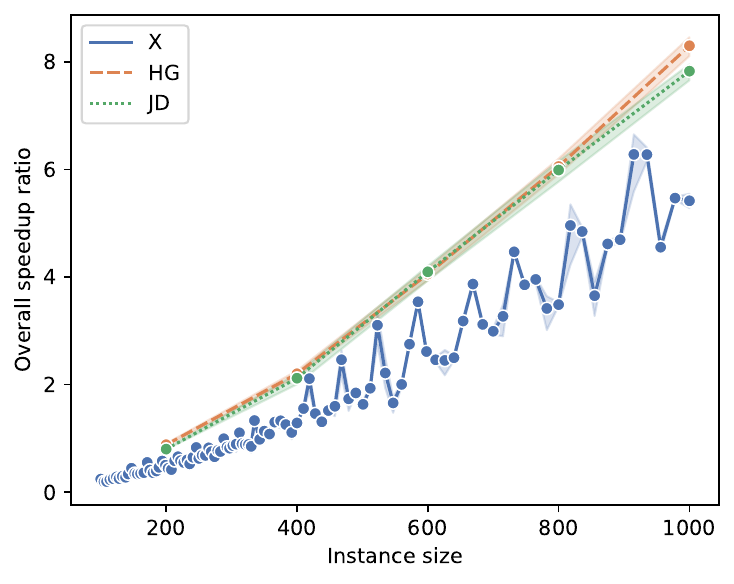}
        \caption{\small Overall speedup of MA-ETGA relative to MA-E.}
        \label{fig:speedup_etga}
    \end{subfigure}
    \caption{Relationship between instance size and speedup ratio across benchmark instances. The x-axis denotes the instance size, while the y-axis reports the speedup of TGA-based algorithms over their CPU-based counterparts.}
    \label{fig:overall_speedup}
\end{figure}

To provide an intuitive assessment of the performance gains achieved by TGA-based algorithms over their traditional CPU-based counterparts, we conducted additional experiments to measure speedup ratios for move evaluation across benchmark instances. Let $t_{\text{CPU}}$ denote the total runtime spent on move evaluations of the CPU-based algorithms (MA-N and MA-E) after 100 iterations, and $t_{\text{TGA}}$ the corresponding runtime for the TGA implementations (MA-NTGA and MA-ETGA) under the same iteration budget. The speedup ratio is defined as $\gamma_s = \frac{t_{\text{CPU}}}{t_{\text{TGA}}}$. Each instance was executed 10 times, with paired algorithm variants (MA-N vs. MA-NTGA and MA-E vs. MA-ETGA) following identical search trajectories in each run to ensure fair and consistent comparisons. Figure~\ref{fig:overall_speedup} reports the resulting speedup ratios across all tested instances, highlighting the relationship between instance size and speedup ratio for different benchmark instances.

The experimental results reveal a clear positive correlation between instance size and speedup ratio across all benchmarks. For the \emph{GH} (VRPTW) and \emph{JD} (VRPSPDTW) instances, both TGA variants exhibit monotonic speedup growth as instance size increases. In particular, MA-NTGA achieves substantial acceleration, with speedup ratios rising from approximately 2 to over 30 when scaling from 200 to 1000 nodes, consistent with theoretical expectations and demonstrating the effectiveness of TGA on large-scale instances. MA-ETGA follows the same trend but with smaller speedup ratios (from about 1 to over 8), as MA-E already reduces move evaluation via neighborhood reduction, leaving less exploitable parallelism for TGA.

A similar trend is observed for the \emph{X} (CVRP) benchmark, where MA-NTGA achieves even higher acceleration on large instances due to the less constrained nature of CVRP, with speedups exceeding 50$\times$ on approximately 1000-node instances. In contrast, no noticeable acceleration is observed for small (around 100-node) instances, where GPU overhead outweighs the performance gains, underscoring that TGA is most beneficial for large and computationally intensive problems. For MA-ETGA on \emph{X} instances, the speedup ratios are substantially lower, as MA-E already eliminates a large portion of move evaluations under the granularity threshold $\theta=20$, leaving limited scope for further acceleration by TGA. Additionally, some variability in speedup is observed among \emph{X} instances, which can be attributed to instance-specific characteristics, particularly the number of vehicles. As discussed in Section~\ref{sec:complexity}, instances requiring more vehicles generate larger scoring tensors, which increases the computational cost to indentify the best move.

\subsection{Acceleration of move evaluation across operators}
\label{sec:analysis_operator}
\begin{figure}[!htbp]
    \centering
    \begin{subfigure}{0.32\textwidth}
        \centering
        \includegraphics[width=\textwidth]{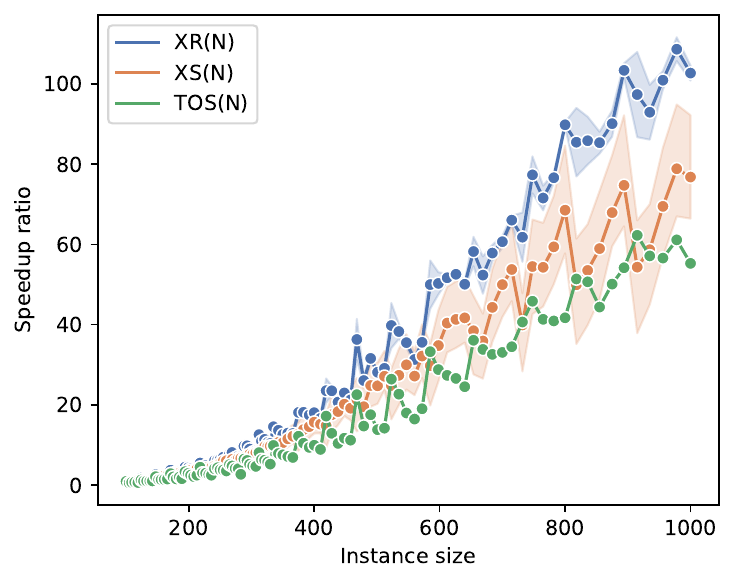}
        \caption{\small Node-based inter-route operators on the \emph{X} instances.}
        \label{fig:x_opt1}
    \end{subfigure}
    \hfill
    \begin{subfigure}{0.32\textwidth}
        \centering
        \includegraphics[width=\textwidth]{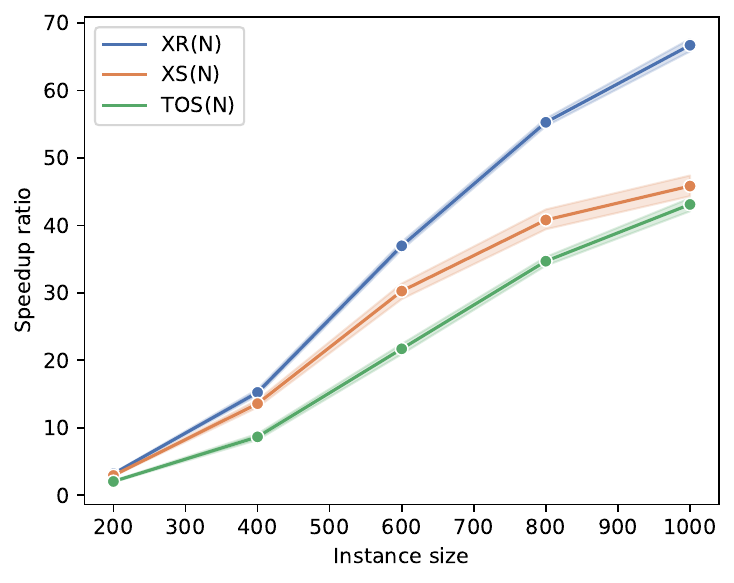}
        \caption{\small Node-based inter-route operators on the \emph{GH} instances.}
        \label{fig:hg_opt1}
    \end{subfigure}
    \hfill
    \begin{subfigure}{0.32\textwidth}
        \centering
        \includegraphics[width=\textwidth]{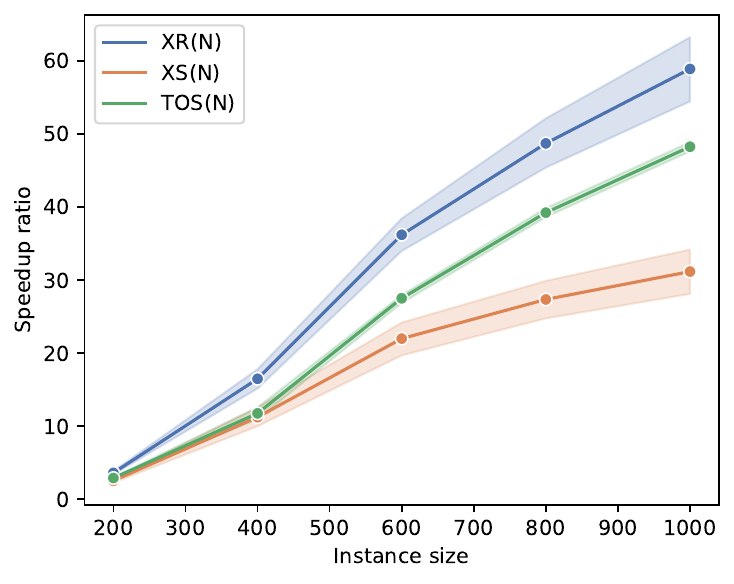}
        \caption{\small Node-based inter-route operators on the \emph{JD} instances.}
        \label{fig:jd_opt1}
    \end{subfigure}
    \hfill
    \begin{subfigure}{0.32\textwidth}
        \centering
        \includegraphics[width=\textwidth]{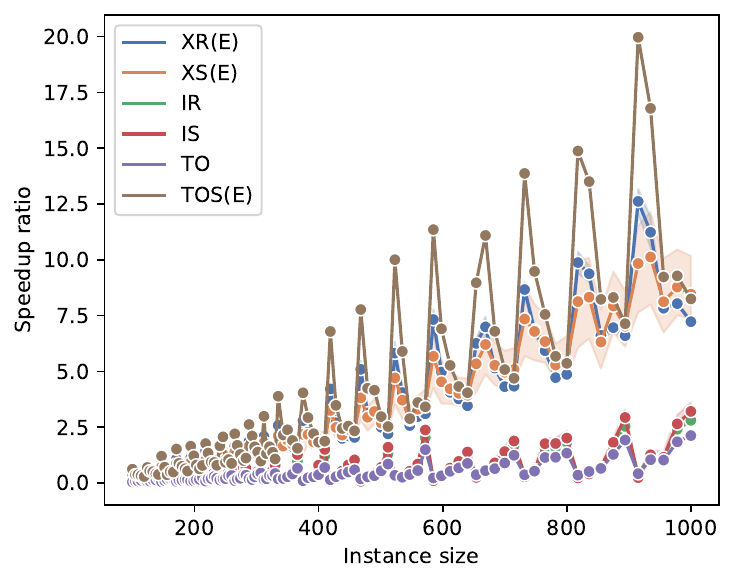}
        \caption{\small intra-route operators and edge-based inter-route operators on the\emph{X} instances.}
        \label{fig:x_opt2}
    \end{subfigure}
    \hfill
    \begin{subfigure}{0.32\textwidth}
        \centering
        \includegraphics[width=\textwidth]{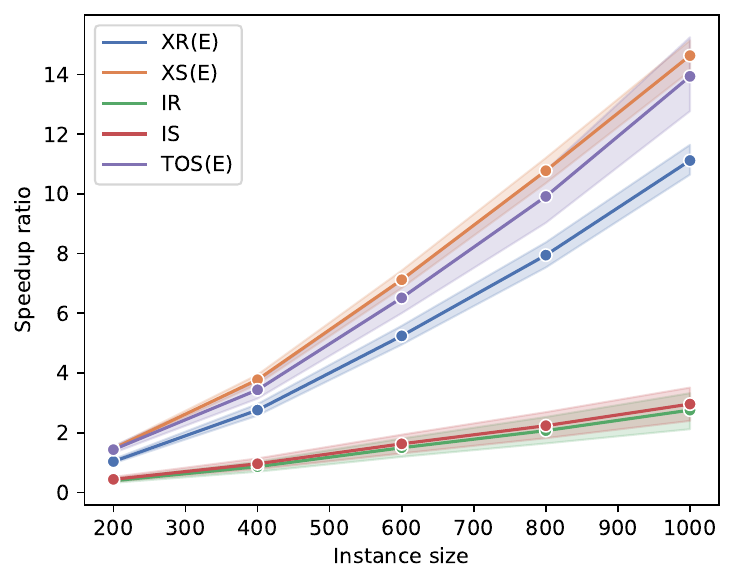}
        \caption{\small intra-route operators and edge-based inter-route operators on the \emph{GH} instances.}
        \label{fig:hg_opt2}
    \end{subfigure}
    \hfill
    \begin{subfigure}{0.32\textwidth}
        \centering
        \includegraphics[width=\textwidth]{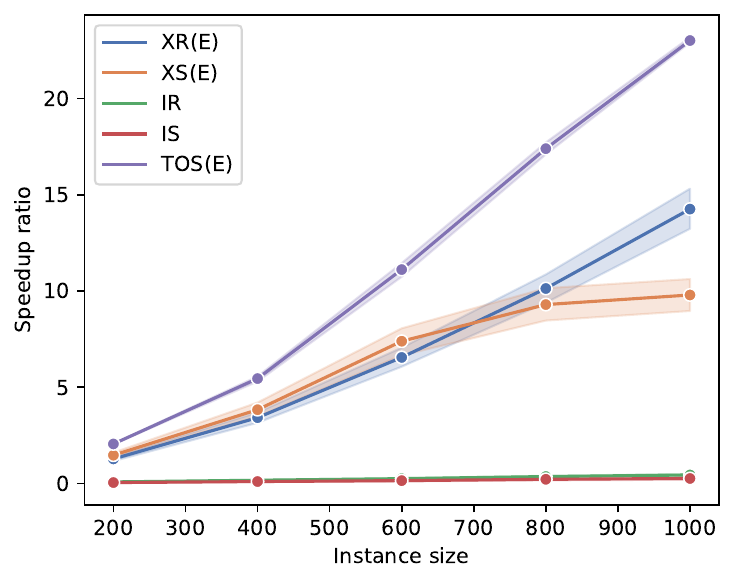}
        \caption{\small intra-route operators and edge-based inter-route operators on the  \emph{JD} instances.}
        \label{fig:jd_opt2}
    \end{subfigure}
    \caption{Speedup ratios of local search operators across benchmark instances, including inter-route operators (\emph{Inter-Relocate}, \emph{Inter-Swap}, \emph{2-opt*}) with node-based (XR(N), XS(N), TOS(N)) and edge-based implementations (XR(E), XS(E), TOS(E)), as well as route-based intra-route operators (\emph{Intra-Relocate} (IR), \emph{Intra-Swap} (IS), \emph{2-opt} (TO)). The x-axis denotes instance size and the y-axis the corresponding speedup ratio.}
    \label{fig:exp_opt}
\end{figure}

We further analyze TGA performance at the operator level by examining speedup gains across benchmark instances of varying sizes. Figure~\ref{fig:exp_opt} shows the speedup ratios for different local search operators, which are calculated as the average runtime of move evaluation for the CPU-based implementation divided by that of the corresponding TGA-based implementation.

Overall, the speedup ratio increases with instance size. TGA achieves particularly significant acceleration for inter-route operators (\emph{Inter-Relocate}, \emph{Inter-Swap}, \emph{2-opt*}), which involve complex cross-route interactions and benefit strongly from GPU parallelization. Node-based implementations reach speedups from 2 up to over 60 for the \emph{GH} and \emph{JD} instances, and exceed 100 for around 1000-node \emph{X} instances. Edge-based implementations also achieve substantial speedups, though slightly lower, as neighborhood reduction strategy already reduces many move evaluations, leaving less room for additional acceleration. In contrast, intra-route operators (\emph{Intra-Relocate}, \emph{Intra-Swap}, \emph{2-opt}) show more modest gains, with speedups up to around 2 for large 1000-node instances. This contrast highlights the particular advantage of the TGA framework for inter-route operators, which require more complex cross-route computations.
\subsection{Time analysis of key steps in TGA}
\label{sec:analysis_step}
\begin{figure}[!htbp]
    \centering
    \begin{subfigure}{0.32\textwidth}
        \centering
        \includegraphics[width=\textwidth]{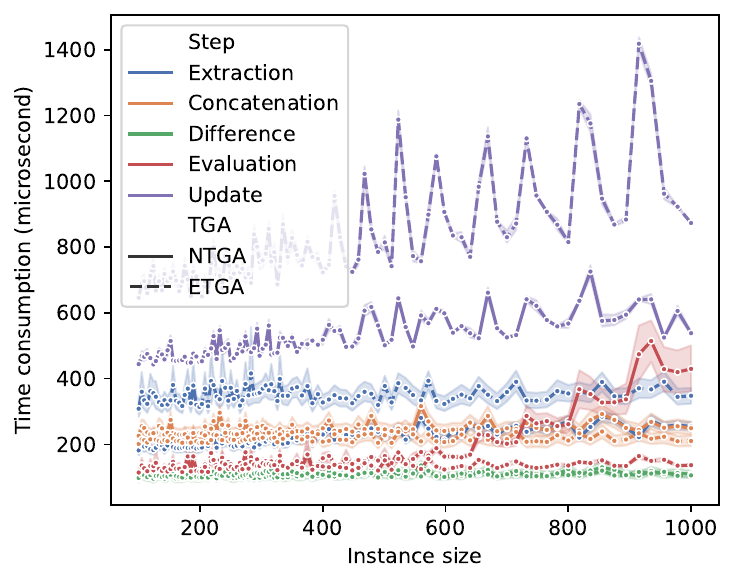}
        \caption{\small Time consumption of TGA's key steps for different instance sizes on \emph{X} instances.}
        \label{fig:exp_vcom_n_t_x}
    \end{subfigure}
    \hfill
    \begin{subfigure}{0.32\textwidth}
        \centering
        \includegraphics[width=\textwidth]{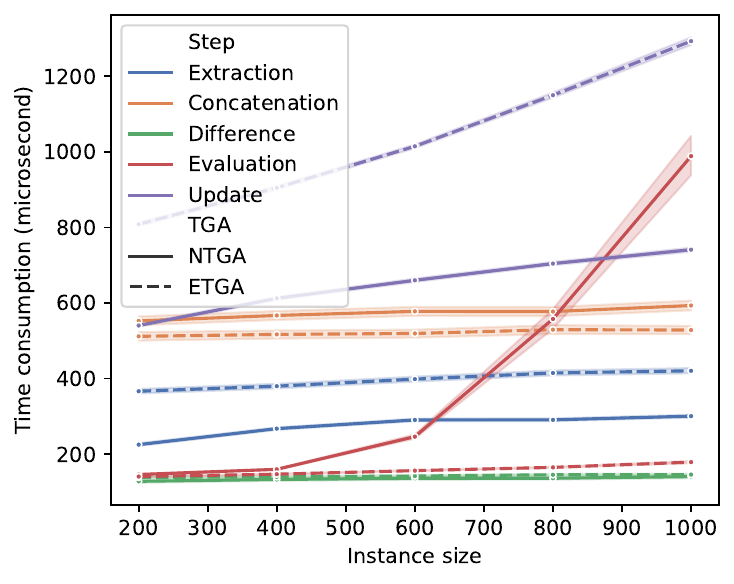}
        \caption{\small Time consumption of TGA's key steps for different instance sizes on \emph{GH} instances.}
        \label{fig:exp_vcom_n_t_hg}
    \end{subfigure}
    \hfill
    \begin{subfigure}{0.32\textwidth}
        \centering
        \includegraphics[width=\textwidth]{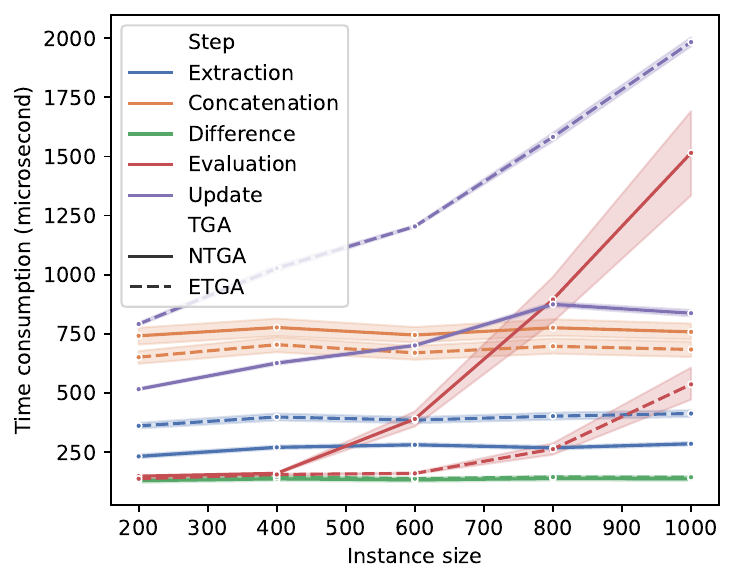}
        \caption{\small Time consumption of TGA's key steps for different instance sizes on \emph{JD} instances.}
        \label{fig:exp_vcom_n_t_jd}
    \end{subfigure}
    \caption{Time analysis of TGA's key steps for varying instance sizes. The x-axis indicates instance size, and the y-axis represents computation time for each step.}
    \label{fig:exp_key_steps} 
\end{figure}

To complement the theoretical complexity analysis in Section~\ref{sec:complexity}, we examine the practical runtime of TGA's key steps (\emph{Extraction}, \emph{Concatenation}, \emph{Differencing}, \emph{Evaluation}, and \emph{Update}) defined in Section~\ref{sec:tensor_operators}. Figure~\ref{fig:exp_key_steps} reports the time consumption of each step across varying instance sizes for both NTGA and ETGA. 

Across all three benchmarks, tensor-based move evaluation steps \emph{Extraction}, \emph{Concatenation}, and \emph{Differencing} show relatively stable runtimes, reflecting near-constant complexity. In contrast, \emph{Evaluation} shows the largest variability, as its cost scales with the size of the \emph{scoring tensor}, which depends on instance characteristics such as the number of customers and vehicles. Therefore, for inter-route operators with large neighborhoods and complex cross-route interactions, \emph{Evaluation} dominates the overall runtime. For intra-route operators, the smaller neighborhoods substantially reduce the \emph{Evaluation} cost, making \emph{Extraction} and \emph{Concatenation} more prominent. Additionally, a clear distinction is observed between NTGA and ETGA in terms of \emph{Evaluation} step. By restricting the effective neighborhood through edge-based implementation, ETGA significantly reduces the \emph{Evaluation} time, particularly on large instances. Finally, the \emph{Update} step emerges as a non-negligible and increasing cost, especially for large instances and in the ETGA. \emph{Update} step involves GPU-CPU synchronization, data transfer, and the update of the solution tensor. In ETGA, this cost is further amplified by the maintenance of \emph{edge index tensors}. Moreover, as ETGA substantially shortens the \emph{Evaluation} step, these fixed overheads account for a larger fraction of the total runtime, making \emph{Update} both absolutely and relatively more expensive.

Overall, this analysis identifies the primary performance bottlenecks of TGA. In NTGA, \emph{Evaluation} dominates due to full neighborhood exploration, whereas in ETGA, the substantial reduction in \emph{Evaluation} time makes \emph{Update} the dominant cost for large-scale instances. These observations suggest potential avenues for performance improvement in future work.

\subsection{Comparative analysis with algorithms in the literature}
Although this study does not aim to achieve new best-known solutions (BKS), we provide a comparative analysis of the proposed TGA-based algorithms (MA-NTGA and MA-ETGA) against leading methods from the literature. For the CVRP and VRPTW, BKS values are taken from the CVRPLIB repository (\url{http://vrp.galgos.inf.puc-rio.br/}) as of December 25, 2025. CVRP comparisons include HGS-2012 \cite{vidal2012hybrid} and HGS-CVRP \cite{vidal2020hybrid}. For the VRPTW, with the objective of minimizing the total travel distance, comparisons include the DIMACS competition reference results and the champion algorithm HGS-DIMACS \cite{kool2022hybrid}. For VRPSPDTW, we report results from the state-of-the-art MA-FIRD method \cite{lei2024memetic}. MA-NTGA and MA-ETGA results correspond to the best solutions in Section~\ref{sec:exp_results}, with all algorithms evaluated under the same termination criteria as reported in their respective references. Although our TGA-based algorithms leverage GPU acceleration while reference algorithms are CPU-based, the CPU specifications provided in the supplementary material indicate that processor performance is broadly comparable.
\begin{figure}[!htbp]
\centering
\includegraphics[width=0.6\textwidth]{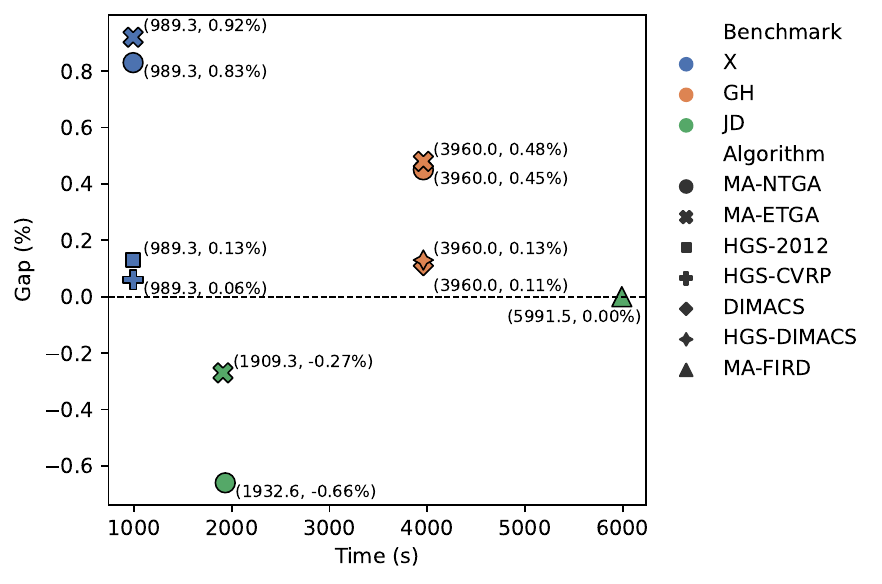}
\caption{Comparative performance of TGA-based algorithms and state-of-the-art methods on benchmark instances. The x-axis shows average computation time (s), and the y-axis shows gap (\%) relative to the BKS.}
\label{fig:sota_plot}
\end{figure}

Figure~\ref{fig:sota_plot} summarizes the results, with detailed numerical results reported in Tables~A8-A12 in the supplementary material. On the \emph{JD} instances, MA-NTGA and MA-ETGA achieve competitive performance, with mean gaps of -0.66\% and -0.27\% relative to BKS in shorter computation times. For the \emph{X} (CVRP) and \emph{GH} (VRPTW) instances, the TGA-based algorithms exhibit larger gaps compared to state-of-the-art methods. Specifically, MA-NTGA attains mean gaps of 0.83\% on \emph{X} and 0.45\% on \emph{GH}, while MA-ETGA achieves 0.92\% and 0.48\% on the same benchmarks, respectively. These results indicate that TGA-based algorithms do not achieve top-tier performance on these two benchmarks, which is expected and can be reasonably explained by several factors.

First, the TGA-based algorithms are built on MA-FIRD, originally designed for the more complex VRPSPDTW rather than for the CVRP or VRPTW. This is particularly impactful for the relatively simpler CVRP, where such sophisticated algorithmic components are not strictly necessary, thereby limiting the competitiveness of TGA-based algorithms on these instances. Second, our algorithms employ a limited set of basic local search operators within the TGA framework, whereas state-of-the-art algorithms typically integrate richer and advanced operators, such as SWAP* \cite{vidal2020hybrid} and RELOCATE* \cite{kool2022hybrid}, which significantly enhance solution quality and search capability. Third, MA-NTGA relies on a best-improvement strategy without neighborhood reduction. Although MA-ETGA introduces neighborhood reduction through an edge-based implementation, it remains more constrained than the sophisticated pruning techniques adopted by state-of-the-art algorithms, which are critical to accelerate the search process and improve search efficiency. Taken together, these factors explain the performance gap observed on the \emph{X} and \emph{GH} instances. 

In summary, while TGA-based algorithms do not yet match leading CPU-based methods, they demonstrate the potential of the TGA framework and GPU-based approaches for vehicle routing problems, providing a promising foundation for future improvements.

\section{Discussion}
\label{sec:discussion}
The TGA framework demonstrates strong extensibility and adaptability. Its attribute-based solution tensor representation allows it to handle a wide range of VRP variants by maintaining problem-specific attribute matrices. Even for more complex VRP variants involving additional decision-making beyond route optimization, such as the multi-depot VRP, the periodic VRP, and the location routing problem, TGA can efficiently optimize the routing component. Its low-coupling architecture fully offloads intensive computations to the GPU, allowing seamless integration into existing local search-based algorithms.

Building on the tensorized implementation of basic local search operators, TGA also supports more advanced operators. For example, \emph{Relocate} and \emph{Swap} can incorporate reversed subsequences by exchanging the first and last node index tensors, as in \emph{2-opt}. Multi-route sequence exchanges can be accommodated by adding tensor dimensions, and filtering strategies, such as tabu-like mechanisms, can be implemented via mask tensors to exclude irrelevant moves.

TGA also exhibits several limitations. Speedups for intra-route operators are less pronounced than for inter-route operators, indicating room for improvement. In node-implemented TGA, the efficiency of the \emph{Evaluation} step is sensitive to the size of the \emph{scoring tensor}. Edge-implemented TGA mitigates this by restricting the neighborhood but introduces additional overhead in the \emph{Update} step due to maintaining \emph{edge index tensors}. These trade-offs suggest that the current solution tensor design does not yet fully support advanced neighborhood reduction techniques commonly used in CPU-based local search. Exploring alternative representations that better accommodate such mechanisms is a promising direction for future work.

Overall, TGA provides a promising direction for leveraging GPU parallelism for solving challenging routing problems. However, comparisons with state-of-the-art algorithms show that it remains early-stage relative to mature CPU-based implementations, with substantial room for improvement in both solution quality and search efficiency. Rather than directly replicating CPU-oriented search strategies which are inherently designed and optimized for CPU architectures, future research should develop search strategies tailored to TGA's parallel execution model. For example, moves are currently evaluated in parallel on the GPU, but acceptance remains sequential, with only one move applied per iteration. In contrast, methods like \emph{Dynasearch} \cite{congram2002iterated} use dynamic programming to identify optimal combinations of multiple independent moves within the neighborhood. Adapting similar ideas could better exploit TGA's massive parallelism to apply multiple moves simultaneously, enhancing both efficiency and solution quality.

Furthermore, the principles of TGA extend beyond local search and can benefit a wide range of optimization methods. Population-based metaheuristics can leverage parallel evaluation and manipulation of multiple solutions on GPUs, while exact methods can exploit tensor representations of partial solutions or subproblems to evaluate feasibility, bounds, and branching decisions in parallel. This approach accelerates traditionally sequential processes such as branch-and-bound or dynamic programming, offering high computational efficiency while maintaining optimality. More broadly, TGA's philosophy can be applied to other combinatorial optimization problems, improving both algorithmic efficiency and performance.

\section{Conclusion}
\label{sec:conclusion}
This work proposes a Tensor-based GPU Acceleration framework, along with two implementations, node-implemented TGA and edge-implemented TGA, for accelerating common local search operators in vehicle routing problems. By leveraging fine-grained GPU parallelism, the proposed framework significantly accelerates neighborhood exploration, which is one of the main computational bottlenecks in local search-based algorithms.

Extensive experiments on three routing problems, CVRP, VRPTW, and VRPSPDTW, confirm the effectiveness and efficiency of our proposed framework, achieving remarkable performance improvements and computational efficiency gains on benchmark instances compared to traditional CPU-based implementations. In-depth analyses of operator-level acceleration and the key steps within the framework valuable insights and reveal critical bottlenecks for future improvement.

Due to its low-coupling architecture and strong extensibility, the TGA framework can be seamlessly integrated into various local search-based routing algorithms, offering a powerful and efficient tool for solving a wide range of vehicle routing problem variants. Beyond vehicle routing, the generality of its attribute-based representation and tensorized evaluation pipeline suggests strong potential for extending the framework to other combinatorial optimization problems and methods. Future work will focus on refining the framework, enhancing GPU utilization, and exploring broader applications within and beyond routing problems.

\section*{Acknowledgments}
\label{sec:acknowledgments}
This work was granted access to the HPC resources of IDRIS under the allocation 2025-AD010617072 made by GENCI. We also acknowledge the HPC resources provided by the Centre de Calcul Intensif des Pays de la Loire (CCIPL), France.

\bibliographystyle{elsarticle-num}
\bibliography{references}

\newpage 
\appendix             

\renewcommand{\thesection}{A\arabic{section}}
\setcounter{section}{0}
\renewcommand{\thetable}{A\arabic{table}}
\setcounter{table}{0}
\renewcommand{\thefigure}{A\arabic{figure}}
\setcounter{figure}{0}

\section*{Supplementary materials of paper ``Speeding up Local Optimization in Vehicle Routing with Tensor-based GPU Acceleration''}  

This document provides supplementary materials for the paper ``Speeding up Local Optimization in Vehicle Routing with Tensor-based GPU Acceleration''. It contains the notation table, the algorithmic framework of the tested algorithms, and detailed computational results for both the proposed algorithm variants and the state-of-the-art comparisons.

\section{Notation}
\label{sec:notation}

Table~\ref{tab:notation} summarizes the key notation used throughout the paper. Note that tensors are uniformly denoted by bold capital letters (e.g., $\mathbf{T}_s$).

\section{Algorithmic framework of the tested algorithms}
\label{sec:algorithm}
This section presents an overview of the algorithms used in the comparative study to evaluate the performance of the proposed tensor-based GPU acceleration (TGA) framework.

The TGA framework is highly extensible and can be integrated into various local search based algorithms. To demonstrate its effectiveness, we incorporate TGA into the MA-FIRD algorithm \cite{lei2024memetic}, which has shown its effectiveness in solving the VRPSPDTW. Since both the CVRP and the VRPTW can be considered special cases of the VRPSPDTW, this integration allows the algorithm to address these variants as well.

MA-FIRD of \cite{lei2024memetic} is a memetic algorithm that combines a feasible and infeasible route descent search (FIRD) and an adaptive route-inheritance crossover (ARIX). The FIRD procedure is a local search procedure using the best-improvement strategy and the sequence concatenation-based move evaluation method. The ARIX crossover generates offspring solutions by inheriting routes from multiple parent solutions and incorporates a learning mechanism to adaptively adjust the number of parents and the heuristics used to complete the partial offspring under construction.

To evaluate the performance of TGA, we modified the initial MA-FIRD algorithm of \cite{lei2024memetic} and incorporated TGA. The resulting general MA framework is illustrated in Algorithm~\ref{algo:framework}. For a given problem instance $\mathcal{I}$, the algorithm begins by generating an initial population $\mathcal{P}$ consisting of $\mu$ candidate solutions (line 4), with the best solution $S^*$ being recorded (line 5). The algorithm then iteratively updates the population by creating new candidate solutions (lines 6-18). In each generation, the adaptive route-inheritance crossover ARIX generates a new solution $S$ (line 7), which is then improved using feasible and infeasible search based on the given local search operators in $\mathcal{O}$ (line 8). Note that the route descent strategy is omitted in this setting to simplify the search process and ensure compatibility with problem variants that do not consider the number of vehicles as an objective. The improved solution $S$ is used to update the population according to the population management strategy (line 9). If an improved best solution is found, $S^*$ is updated, and the stagnation counter is reset to zero (lines 11-13). Otherwise, the stagnation generation counter $\varphi_{st}$ increments (lines 14-15). The algorithm terminates and returns the best solution $S^*$ (line 19) when the given termination condition is reached (line 6).

\begin{algorithm}[!ht] 
    \caption{General MA framework}
    \label{algo:framework} 
    \begin{algorithmic}[1]
        \STATE \textbf{Input}: Instance $\mathcal{I}$, population size $\mu$, set of operators $\mathcal{O} = \{O_1, O_2, \dots, O_n\}$.
        \STATE \textbf{Output}: The best solution $S^*$.
        \STATE $\varphi \gets 0, \varphi_{st} \gets 0$ \hfill /* Current generation and stagnation generation counter */
        \STATE $\mathcal{P} \gets Initialization(\mathcal{I}, \mu)$ 
        \STATE $S^* \gets BestSolution(\mathcal{P})$ \hfill /* Record current best solution */
        \WHILE {The given termination condition is not met}
            \STATE $S \gets ARIX(\mathcal{P})$ 
            \STATE $S \gets FeasibleInfeasibleSearch(S, \mathcal{O})$ 
            \STATE $\mathcal{P} \gets UpdatePopulation(S, \mathcal{P})$ 
            \STATE $S_{best} \gets BestSolution(\mathcal{P})$ \hfill /* The best solution in this generation */
            \IF {$f(S_{best}) < f(S^*)$}
                \STATE $S^* \gets S_{best}$ \hfill /* Update the best solution */
                \STATE $\varphi_{st} \gets 0$
            \ELSE
                \STATE $\varphi_{st} \gets \varphi_{st} + 1$ \hfill /* Stagnation counter is incremented */
            \ENDIF
            \STATE $\varphi = \varphi + 1$ 
        \ENDWHILE
        \STATE \textbf{return} $S^*$ 
    \end{algorithmic}
\end{algorithm}

\begin{algorithm}[!ht]
    \caption{Feasible and infeasible search}
    \label{algo:FIS} 
    \begin{algorithmic}[1]
        \STATE \textbf{Input}: Input solution $S$, set of operators $\mathcal{O} = \{O_1, O_2, \dots, O_n\}$.
        \STATE \textbf{Output}: Improved solution $S$.
        \STATE Initialize tensor $\mathbf{T}_s$ on GPU based on the solution $S$. 
        \WHILE {The given termination criterion is not met}
            \STATE Select an operator $O_i \in \mathcal{O}$.
            \STATE $\zeta \gets Evaluate(\mathbf{T}_s, S, O_i)$ \hfill /*Evaluate the neighborhood of the operator $O_i$ and obtain the best move $\zeta$*/.
            \IF {The move $\zeta$ is accepted based on the defined evaluation}
                \STATE Update current solution $S$ based on the move $\zeta$.
                \STATE Update tensor $\mathbf{T}_s$ based on current solution $S$.
            \ENDIF
        \ENDWHILE 
        \STATE \textbf{return} $S$ 
    \end{algorithmic}
\end{algorithm}
Within the MA framework, the feasible and infeasible search constitutes the core local search component responsible for search intensification and distinguishes the four algorithmic variants considered in our comparative study. Algorithm~\ref{algo:FIS} presents the general procedure of the feasible and infeasible search for the TGA-based algorithms (MA-NTGA and MA-ETGA), in which the proposed tensor-based GPU acceleration is employed for move evaluation (line 6). In contrast, their CPU-based counterparts (MA-N and MA-E) exclude lines 3 and 9 of Algorithm~\ref{algo:FIS} and rely on traditional CPU implementations of the local search operators for move evaluation (line 6). Both the TGA-based and CPU-based algorithms adopt the same best-improvement local search strategy. MA-N and MA-NTGA evaluate the full neighborhood without reduction, whereas MA-E and MA-ETGA apply neighborhood reduction.

\section{Detailed computational results}
\label{sec:results}

Tables~\ref{tab:exp_x}-\ref{tab:exp_jd} present detailed computational results based on 10 runs for four algorithm variants, MA-N, MA-E, MA-NTGA, and MA-ETGA, which are evaluated on three benchmark sets: 100 \emph{X} (CVRP) instances \cite{uchoa2017new}, 90 selected \emph{GH} (VRPTW) instances \cite{gehring1999parallel}, and 20 \emph{JD} (VRPSPDTW) instances \cite{liu2021memetic}.

MA-N and MA-E are CPU-based algorithms employing traditional local search evaluation methods, whereas MA-NTGA and MA-ETGA integrate the proposed tensor-based GPU acceleration (TGA). MA-N serves as the baseline algorithm for performance comparison.

Each table reports the instance name (\emph{Instance}), the number of customers ($N_C$), the average number of vehicles ($M$), the best objective value ($f^*$), and the average objective value ($f$) over 10 independent runs for each algorithm variant. The column $\text{``Gap(\%)''}$ indicates the percentage improvement in average objective value relative to MA-N, where a negative value corresponds to better performance.

For the \emph{X} (CVRP) and \emph{GH} (VRPTW) instances, the unified termination conditions with time limits from the literature \cite{vidal2012hybrid,vidal2020hybrid,kool2022hybrid} are adopted, and the corresponding limits are reported in column $t$ (in seconds). For the \emph{JD} (VRPSPDTW) instances, compound termination conditions are applied following \cite{lei2024memetic}, and the actual average computation times are reported in column $t$.

\section{Detailed computational results for the state-of-the-art comparisons}
\label{sec:sota_results}
This section provides detailed comparative results between the proposed TGA-based algorithms (MA-NTGA and MA-ETGA) and state-of-the-art methods on three benchmark sets: \emph{X} (CVRP), \emph{GH} (VRPTW), and \emph{JD} (VRPSPDTW). Although our TGA-based algorithms leverage GPU acceleration while reference algorithms are CPU-based, Table~\ref{tab:cpu} reports PassMark CPU benchmark scores (\url{https://www.passmark.com/}) for both our experimental CPU and those used in the reference studies. The reported differences show that our CPU performance is within $\pm10\%$ of the references, indicating broadly comparable CPU performance.
\begin{table}[!ht]
\centering
\caption{Processor specifications and corresponding CPU benchmark scores.}
\label{tab:cpu}
\resizebox{\textwidth}{!}{%
\begin{tabular}{llccc}
\hline
Algorithm & Processor & Base frequency & CPU mark & Difference (\%) \\ \hline
HGS-2012 & Intel Xeon Gold 6148 & 2.40\,GHz & 2072 & -8.2\% \\
HGS-CVRP & Intel Xeon Gold 6148 & 2.40\,GHz & 2072 & -8.2\% \\
HGS-DIMACS & Intel Core i7-6850K & 3.60\,GHz & 2388 & +5.8\% \\
MA-FIRD & Intel Xeon E5-2690 v4 & 2.60\,GHz & 2069 & -8.3\% \\
Ours & Intel Xeon Gold 6248 & 2.50\,GHz & 2257 & 0\% \\ \hline
\end{tabular}
}
\end{table}

Tables~\ref{tab:sota_x}-\ref{tab:sota_jd} show detailed comparative results between our TGA-based algorithms (MA-NTGA and MA-ETGA) and state-of-the-art methods on the three benchmark sets, respectively. For the CVRP and VRPTW, the BKS values are obtained from the CVRPLIB repository (\url{http://vrp.galgos.inf.puc-rio.br/}) as of December 25, 2025. For the CVRP, we use results from HGS-2012 \cite{vidal2012hybrid} and HGS-CVRP \cite{vidal2020hybrid}. For the VRPTW, with the objective of minimizing the total travel distance, we report results from the DIMACS competition, including both the official DIMACS reference results and the champion team's algorithm, HGS-DIMACS \cite{kool2022hybrid}. For the VRPSPDTW, we report the best results from the state-of-the-art MA-FIRD method \cite{lei2024memetic}. Our MA-NTGA and MA-ETGA results are derived from the best solutions obtained in the experiments reported in Section~\ref{sec:results}.

Each table reports the instance name (\emph{Instance}), the number of customers ($N_C$), the best-known solution (BKS), the best objective value ($f^*$) achieved by each reference algorithm, and the corresponding gap ($\text{Gap(\%)}$) relative to the BKS. The time column ($t$) follows the same definition and settings as in Section~\ref{sec:results}.

\input{tables/notation.tex}

\input{tables/exp_x.tex}
\input{tables/exp_hg.tex}
\input{tables/exp_jd.tex}

\input{tables/sota_x.tex}
\input{tables/sota_hg.tex}
\input{tables/sota_jd.tex}

\end{document}

%% file: tables/notation.tex
\begin{table}[!ht]
\centering
\caption{Summary of key notation.}
\label{tab:notation}
\scriptsize
\renewcommand{\arraystretch}{0.95}
\begin{tabular}{@{}ll@{}}
\hline
Notation & Description \\
\hline
\multicolumn{2}{l}{\textbf{Problem Description}} \\
$\mathcal{G}=(\mathcal{V}, \mathcal{E})$ & Complete graph with nodes $\mathcal{V}$ and edges $\mathcal{E}$ \\
$\mathcal{V}=\{v_0, v_1, \dots, v_{N_C}\}$ & Nodes: depot $v_0$ and customers $\{v_1,\dots,v_{N_C}\}$ \\
$\mathcal{E}=\{e_{ij} \mid v_i, v_j \in \mathcal{V}\}$ & Set of edges connecting nodes $v_i$ and $v_j$ \\
$N_C$ & Number of customers \\
$d_i, p_i$ & Delivery and pickup demands of node $v_i$ \\
$e_i, l_i, s_i$ & Earliest time, latest time, and service time of node $v_i$ \\
\hline
\end{tabular}
\end{table}

%% file: tables/exp_x.tex
\begin{table}[!ht]
\centering
\caption{Comparative results on the 100 \textit{X} benchmark instances for MA-N, MA-E, MA-NTGA, and MA-ETGA.}
\label{tab:exp_x}
\scriptsize
\setlength{\tabcolsep}{2.0pt}
\renewcommand{\arraystretch}{0.88}
\resizebox{\linewidth}{!}{%
\begin{tabular}{lrrrrrrrrrrrrrrrrr}
\hline
\multirow{2}{*}{Instance} & \multicolumn{1}{c}{\multirow{2}{*}{$N_C$}} & \multicolumn{1}{c}{\multirow{2}{*}{$t$}} & \multicolumn{3}{c}{MA-N} & \multicolumn{4}{c}{MA-E} & \multicolumn{4}{c}{MA-NTGA} & \multicolumn{4}{c}{MA-ETGA} \\
\cmidrule(l){4-6} \cmidrule(l){7-10} \cmidrule(l){11-14} \cmidrule(l){15-18}
    & \multicolumn{1}{c}{} & \multicolumn{1}{c}{} & \multicolumn{1}{c}{$M$} & \multicolumn{1}{c}{$f^*$} & \multicolumn{1}{c}{$f$} & \multicolumn{1}{c}{$M$} & \multicolumn{1}{c}{$f^*$} & \multicolumn{1}{c}{$f$} & \multicolumn{1}{c}{Gap(\%)} & \multicolumn{1}{c}{$M$} & \multicolumn{1}{c}{$f^*$} & \multicolumn{1}{c}{$f$} & \multicolumn{1}{c}{Gap(\%)} & \multicolumn{1}{c}{$M$} & \multicolumn{1}{c}{$f^*$} & \multicolumn{1}{c}{$f$} & \multicolumn{1}{c}{Gap(\%)} \\
\hline
X-n101-k25 & 100 & 240.0 & 26.0 & 27591.0 & 27597.8 & 26.0 & 27591.0 & 27603.4 & 0.02 & 26.0 & 27591.0 & 27605.0 & 0.03 & 26.0 & 27591.0 & 27599.8 & 0.01 \\
X-n106-k14 & 105 & 252.0 & 14.0 & 26381.0 & 26481.4 & 14.0 & 26392.0 & 26435.4 & -0.17 & 14.0 & 26403.0 & 26464.2 & -0.06 & 14.0 & 26403.0 & 26464.6 & -0.06 \\
X-n110-k13 & 109 & 261.6 & 13.0 & 14971.0 & 14988.9 & 13.0 & 14971.0 & 14992.6 & 0.02 & 13.0 & 14971.0 & 14999.8 & 0.07 & 13.0 & 14971.0 & 15001.4 & 0.08 \\
X-n115-k10 & 114 & 273.6 & 10.0 & 12747.0 & 12758.1 & 10.0 & 12747.0 & 12773.9 & 0.12 & 10.0 & 12747.0 & 12751.7 & -0.05 & 10.0 & 12747.0 & 12754.8 & -0.03 \\
X-n120-k6 & 119 & 285.6 & 6.0 & 13332.0 & 13397.8 & 6.0 & 13332.0 & 13380.5 & -0.13 & 6.0 & 13332.0 & 13373.8 & -0.18 & 6.0 & 13332.0 & 13363.6 & -0.26 \\
X-n125-k30 & 124 & 297.6 & 30.0 & 55562.0 & 55641.0 & 30.0 & 55551.0 & 55636.3 & -0.01 & 30.0 & 55626.0 & 55721.0 & 0.14 & 30.0 & 55600.0 & 55682.4 & 0.07 \\
\hline
\end{tabular}%
}
\end{table}

\begin{table}[!ht]
\centering
\caption{Comparative results on the 100 \textit{X} benchmark instances for MA-N, MA-E, MA-NTGA, and MA-ETGA (continued).}
\label{tab:exp_x_continued}
\scriptsize
\setlength{\tabcolsep}{2.0pt}
\renewcommand{\arraystretch}{0.88}
\resizebox{\linewidth}{!}{%
\begin{tabular}{lrrrrrrrrrrrrrrrrr}
\hline
\multirow{2}{*}{Instance} & \multicolumn{1}{c}{\multirow{2}{*}{$N_C$}} & \multicolumn{1}{c}{\multirow{2}{*}{$t$}} & \multicolumn{3}{c}{MA-N} & \multicolumn{4}{c}{MA-E} & \multicolumn{4}{c}{MA-NTGA} & \multicolumn{4}{c}{MA-ETGA} \\
\cmidrule(l){4-6} \cmidrule(l){7-10} \cmidrule(l){11-14} \cmidrule(l){15-18}
    & \multicolumn{1}{c}{} & \multicolumn{1}{c}{} & \multicolumn{1}{c}{$M$} & \multicolumn{1}{c}{$f^*$} & \multicolumn{1}{c}{$f$} & \multicolumn{1}{c}{$M$} & \multicolumn{1}{c}{$f^*$} & \multicolumn{1}{c}{$f$} & \multicolumn{1}{c}{Gap(\%)} & \multicolumn{1}{c}{$M$} & \multicolumn{1}{c}{$f^*$} & \multicolumn{1}{c}{$f$} & \multicolumn{1}{c}{Gap(\%)} & \multicolumn{1}{c}{$M$} & \multicolumn{1}{c}{$f^*$} & \multicolumn{1}{c}{$f$} & \multicolumn{1}{c}{Gap(\%)} \\
\hline
X-n613-k62 & 612 & 1468.8 & 62.0 & 62921.0 & 63335.5 & 62.0 & 61330.0 & 62086.2 & -1.97 & 62.0 & 60571.0 & 60698.4 & -4.16 & 62.0 & 60687.0 & 60861.2 & -3.91 \\
X-n627-k43 & 626 & 1502.4 & 43.5 & 64169.0 & 64712.4 & 43.3 & 63621.0 & 64700.9 & -0.02 & 43.0 & 63015.0 & 63320.0 & -2.15 & 43.0 & 63096.0 & 63305.3 & -2.17 \\
X-n641-k35 & 640 & 1536.0 & 35.0 & 65633.0 & 65826.2 & 35.0 & 64905.0 & 65975.3 & 0.23 & 35.0 & 64516.0 & 64847.8 & -1.49 & 35.0 & 64684.0 & 64848.6 & -1.49 \\
X-n655-k131 & 654 & 1569.6 & 131.0 & 107273.0 & 107394.4 & 131.0 & 107343.0 & 107908.7 & 0.48 & 131.0 & 107248.0 & 107338.8 & -0.05 & 131.0 & 107269.0 & 107354.4 & -0.04 \\
X-n670-k130 & 669 & 1605.6 & 133.3 & 149871.0 & 150288.7 & 133.7 & 148928.0 & 149451.6 & -0.56 & 133.0 & 147491.0 & 147731.1 & -1.70 & 134.8 & 147803.0 & 148602.4 & -1.12 \\
X-n685-k75 & 684 & 1641.6 & 75.2 & 72087.0 & 72821.3 & 75.6 & 70656.0 & 71193.7 & -2.24 & 75.6 & 69319.0 & 69634.2 & -4.38 & 75.9 & 69510.0 & 69859.7 & -4.07 \\
\hline
\end{tabular}%
}
\end{table}

%% file: tables/exp_hg.tex
\begin{table}[!ht]
\centering
\caption{Comparative results on the 90 \textit{GH} benchmark instances for MA-N, MA-E, MA-NTGA, and MA-ETGA.}
\label{tab:exp_hg}
\scriptsize
\setlength{\tabcolsep}{2.0pt}
\renewcommand{\arraystretch}{0.88}
\resizebox{\linewidth}{!}{%
\begin{tabular}{lrrrrrrrrrrrrrrrrr}
\hline
\multirow{2}{*}{Instance} & \multicolumn{1}{c}{\multirow{2}{*}{$N_C$}} & \multicolumn{1}{c}{\multirow{2}{*}{$t$}} & \multicolumn{3}{c}{MA-N} & \multicolumn{4}{c}{MA-E} & \multicolumn{4}{c}{MA-NTGA} & \multicolumn{4}{c}{MA-ETGA} \\
\cmidrule(l){4-6} \cmidrule(l){7-10} \cmidrule(l){11-14} \cmidrule(l){15-18}
    & \multicolumn{1}{c}{} & \multicolumn{1}{c}{} & \multicolumn{1}{c}{$M$} & \multicolumn{1}{c}{$f^*$} & \multicolumn{1}{c}{$f$} & \multicolumn{1}{c}{$M$} & \multicolumn{1}{c}{$f^*$} & \multicolumn{1}{c}{$f$} & \multicolumn{1}{c}{Gap(\%)} & \multicolumn{1}{c}{$M$} & \multicolumn{1}{c}{$f^*$} & \multicolumn{1}{c}{$f$} & \multicolumn{1}{c}{Gap(\%)} & \multicolumn{1}{c}{$M$} & \multicolumn{1}{c}{$f^*$} & \multicolumn{1}{c}{$f$} & \multicolumn{1}{c}{Gap(\%)} \\
\hline
C1\_2\_1 & 200 & 1800.0 & 20.0 & 2698.6 & 2698.6 & 20.0 & 2698.6 & 2698.6 & 0.00 & 20.0 & 2698.6 & 2698.6 & 0.00 & 20.0 & 2698.6 & 2698.6 & 0.00 \\
C1\_2\_5 & 200 & 1800.0 & 20.0 & 2694.9 & 2694.9 & 20.0 & 2694.9 & 2694.9 & 0.00 & 20.0 & 2694.9 & 2694.9 & 0.00 & 20.0 & 2694.9 & 2694.9 & 0.00 \\
C1\_2\_10 & 200 & 1800.0 & 19.0 & 2624.7 & 2625.6 & 19.0 & 2624.7 & 2626.5 & 0.04 & 19.0 & 2624.7 & 2625.6 & 0.00 & 19.0 & 2624.7 & 2625.5 & 0.00 \\
C2\_2\_1 & 200 & 1800.0 & 6.9 & 1922.1 & 1922.3 & 7.0 & 1922.1 & 1922.1 & -0.01 & 7.0 & 1922.1 & 1922.3 & 0.00 & 7.0 & 1922.1 & 1922.1 & -0.01 \\
C2\_2\_5 & 200 & 1800.0 & 7.0 & 1869.6 & 1869.6 & 7.0 & 1869.6 & 1869.6 & 0.00 & 6.9 & 1869.6 & 1869.9 & 0.01 & 7.0 & 1869.6 & 1869.6 & 0.00 \\
C2\_2\_10 & 200 & 1800.0 & 6.9 & 1791.2 & 1793.8 & 7.0 & 1791.2 & 1792.0 & -0.10 & 6.9 & 1791.2 & 1792.6 & -0.07 & 7.0 & 1791.2 & 1792.1 & -0.09 \\
\hline
\end{tabular}%
}
\end{table}

\begin{table}[!ht]
\centering
\caption{Comparative results on the 90 \textit{GH} benchmark instances for MA-N, MA-E, MA-NTGA, and MA-ETGA (continued).}
\label{tab:exp_hg_continued}
\scriptsize
\setlength{\tabcolsep}{2.0pt}
\renewcommand{\arraystretch}{0.88}
\resizebox{\linewidth}{!}{%
\begin{tabular}{lrrrrrrrrrrrrrrrrr}
\hline
\multirow{2}{*}{Instance} & \multicolumn{1}{c}{\multirow{2}{*}{$N_C$}} & \multicolumn{1}{c}{\multirow{2}{*}{$t$}} & \multicolumn{3}{c}{MA-N} & \multicolumn{4}{c}{MA-E} & \multicolumn{4}{c}{MA-NTGA} & \multicolumn{4}{c}{MA-ETGA} \\
\cmidrule(l){4-6} \cmidrule(l){7-10} \cmidrule(l){11-14} \cmidrule(l){15-18}
    & \multicolumn{1}{c}{} & \multicolumn{1}{c}{} & \multicolumn{1}{c}{$M$} & \multicolumn{1}{c}{$f^*$} & \multicolumn{1}{c}{$f$} & \multicolumn{1}{c}{$M$} & \multicolumn{1}{c}{$f^*$} & \multicolumn{1}{c}{$f$} & \multicolumn{1}{c}{Gap(\%)} & \multicolumn{1}{c}{$M$} & \multicolumn{1}{c}{$f^*$} & \multicolumn{1}{c}{$f$} & \multicolumn{1}{c}{Gap(\%)} & \multicolumn{1}{c}{$M$} & \multicolumn{1}{c}{$f^*$} & \multicolumn{1}{c}{$f$} & \multicolumn{1}{c}{Gap(\%)} \\
\hline
R1\_6\_1 & 600 & 3600.0 & 61.2 & 21711.0 & 21858.9 & 59.1 & 21377.2 & 21515.9 & -1.57 & 58.1 & 21314.5 & 21352.3 & -2.32 & 58.2 & 21315.5 & 21385.5 & -2.17 \\
R1\_6\_5 & 600 & 3600.0 & 56.6 & 19884.8 & 19992.1 & 56.2 & 19552.0 & 19641.4 & -1.75 & 56.0 & 19435.8 & 19523.9 & -2.34 & 56.0 & 19463.8 & 19545.8 & -2.23 \\
R1\_6\_10 & 600 & 3600.0 & 55.2 & 18226.7 & 18355.5 & 55.0 & 17958.7 & 18040.4 & -1.72 & 55.0 & 17711.0 & 17799.5 & -3.03 & 55.0 & 17795.1 & 17861.1 & -2.69 \\
R2\_6\_1 & 600 & 3600.0 & 20.0 & 15654.4 & 16012.2 & 21.4 & 15259.8 & 15595.6 & -2.60 & 21.4 & 15274.1 & 15693.4 & -1.99 & 23.4 & 15194.9 & 15418.9 & -3.71 \\
R2\_6\_5 & 600 & 3600.0 & 15.7 & 14519.5 & 14786.5 & 17.3 & 13962.7 & 14226.6 & -3.79 & 18.2 & 13895.0 & 14120.2 & -4.51 & 18.3 & 13926.5 & 14150.4 & -4.30 \\
R2\_6\_10 & 600 & 3600.0 & 13.0 & 12465.8 & 12633.6 & 13.1 & 12024.2 & 12245.4 & -3.07 & 13.8 & 12010.4 & 12084.8 & -4.34 & 13.6 & 11988.6 & 12090.3 & -4.30 \\
\hline
\end{tabular}%
}
\end{table}

%% file: tables/exp_jd.tex
\begin{table}[!ht]
\centering
\caption{Comparative results on the 20 \textit{JD} benchmark instances for MA-N, MA-E, MA-NTGA, and MA-ETGA.}
\label{tab:exp_jd}
\scriptsize
\setlength{\tabcolsep}{2.0pt}
\renewcommand{\arraystretch}{0.88}
\resizebox{\linewidth}{!}{%
\begin{tabular}{lrrrrrrrrrrrrrrrrrrrr}
\hline
\multirow{2}{*}{Instance} & \multicolumn{1}{c}{\multirow{2}{*}{$N_C$}} & \multicolumn{4}{c}{MA-N} & \multicolumn{5}{c}{MA-E} & \multicolumn{5}{c}{MA-NTGA} & \multicolumn{5}{c}{MA-ETGA} \\
\cmidrule(l){3-6} \cmidrule(l){7-11} \cmidrule(l){12-16} \cmidrule(l){17-21}
    & \multicolumn{1}{c}{} & \multicolumn{1}{c}{$M$} & \multicolumn{1}{c}{$f^*$} & \multicolumn{1}{c}{$f$} & \multicolumn{1}{c}{$t$} & \multicolumn{1}{c}{$M$} & \multicolumn{1}{c}{$f^*$} & \multicolumn{1}{c}{$f$} & \multicolumn{1}{c}{$t$} & \multicolumn{1}{c}{Gap(\%)} & \multicolumn{1}{c}{$M$} & \multicolumn{1}{c}{$f^*$} & \multicolumn{1}{c}{$f$} & \multicolumn{1}{c}{$t$} & \multicolumn{1}{c}{Gap(\%)} & \multicolumn{1}{c}{$M$} & \multicolumn{1}{c}{$f^*$} & \multicolumn{1}{c}{$f$} & \multicolumn{1}{c}{$t$} & \multicolumn{1}{c}{Gap(\%)} \\
\hline
F201 & 200 & 41.0 & 64458.1 & 64690.3 & 908.5 & 41.0 & 64437.6 & 64705.0 & 439.5 & 0.02 & 41.0 & 64297.8 & 64550.4 & 424.0 & -0.22 & 41.1 & 64483.1 & 64812.3 & 516.1 & 0.19 \\
F202 & 200 & 42.8 & 64687.0 & 64944.5 & 1089.2 & 42.9 & 64848.5 & 65002.1 & 376.6 & 0.09 & 42.9 & 64758.0 & 64875.4 & 415.1 & -0.11 & 42.8 & 64698.3 & 64969.9 & 608.1 & 0.04 \\
F203 & 200 & 41.2 & 65667.3 & 65994.2 & 1067.7 & 41.2 & 65619.3 & 66042.1 & 511.8 & 0.07 & 41.2 & 65526.9 & 65967.3 & 495.1 & -0.04 & 41.0 & 65664.1 & 65854.0 & 614.9 & -0.21 \\
F204 & 200 & 41.7 & 64309.7 & 64755.8 & 1164.3 & 41.6 & 64461.7 & 64780.3 & 442.2 & 0.04 & 41.5 & 64179.1 & 64629.0 & 532.0 & -0.20 & 41.8 & 64310.7 & 64835.5 & 641.4 & 0.12 \\
F401 & 400 & 76.4 & 117113.0 & 117937.6 & 6489.8 & 76.6 & 117068.0 & 117779.9 & 2422.3 & -0.13 & 76.4 & 116999.0 & 117512.5 & 841.8 & -0.36 & 76.8 & 117057.0 & 118032.7 & 1056.2 & 0.08 \\
F402 & 400 & 80.2 & 122364.0 & 123627.1 & 6798.0 & 79.8 & 122606.0 & 123308.8 & 2674.6 & -0.26 & 80.0 & 122429.0 & 123059.2 & 865.5 & -0.46 & 79.9 & 122369.0 & 123023.9 & 1567.1 & -0.49 \\
\hline
\end{tabular}%
}
\end{table}

%% file: tables/sota_x.tex
\begin{table}[!ht]
\centering
\caption{Comparative results on the 100 \textit{X} benchmark instances for HGS-2012, HGS-CVRP, MA-NTGA, and MA-ETGA.}
\label{tab:sota_x}
\scriptsize
\renewcommand{\arraystretch}{0.68}
\resizebox{\linewidth}{!}{%
\begin{tabular}{lrrrrrrrrrrr}
\hline
\multirow{2}{*}{Instance} & \multicolumn{1}{c}{\multirow{2}{*}{$N_C$}} & \multicolumn{1}{c}{\multirow{2}{*}{BKS}} & \multicolumn{1}{c}{\multirow{2}{*}{$t$}} & \multicolumn{2}{c}{HGS-2012} & \multicolumn{2}{c}{HGS-CVRP} & \multicolumn{2}{c}{MA-NTGA} & \multicolumn{2}{c}{MA-ETGA} \\
\cmidrule(l){5-6} \cmidrule(l){7-8} \cmidrule(l){9-10} \cmidrule(l){11-12}
    & \multicolumn{1}{c}{} & \multicolumn{1}{c}{} & \multicolumn{1}{c}{} & \multicolumn{1}{c}{$f^*$} & \multicolumn{1}{c}{Gap(\%)} & \multicolumn{1}{c}{$f^*$} & \multicolumn{1}{c}{Gap(\%)} & \multicolumn{1}{c}{$f^*$} & \multicolumn{1}{c}{Gap(\%)} & \multicolumn{1}{c}{$f^*$} & \multicolumn{1}{c}{Gap(\%)} \\
\hline
X-n101-k25 & 100 & 27591.0 & 240.0 & 27591.0 & 0.00 & 27591.0 & 0.00 & 27591.0 & 0.00 & 27591.0 & 0.00 \\
X-n106-k14 & 105 & 26362.0 & 252.0 & 26387.0 & 0.09 & 26364.0 & 0.01 & 26403.0 & 0.16 & 26403.0 & 0.16 \\
X-n110-k13 & 109 & 14971.0 & 261.6 & 14971.0 & 0.00 & 14971.0 & 0.00 & 14971.0 & 0.00 & 14971.0 & 0.00 \\
X-n115-k10 & 114 & 12747.0 & 273.6 & 12747.0 & 0.00 & 12747.0 & 0.00 & 12747.0 & 0.00 & 12747.0 & 0.00 \\
X-n120-k6 & 119 & 13332.0 & 285.6 & 13332.0 & 0.00 & 13332.0 & 0.00 & 13332.0 & 0.00 & 13332.0 & 0.00 \\
X-n125-k30 & 124 & 55539.0 & 297.6 & 55539.0 & 0.00 & 55539.0 & 0.00 & 55626.0 & 0.16 & 55600.0 & 0.11 \\
\hline
\end{tabular}%
}
\end{table}

\begin{table}[!ht]
\centering
\caption{Comparative results on the 100 \textit{X} benchmark instances for HGS-2012, HGS-CVRP, MA-NTGA, and MA-ETGA (continued).}
\label{tab:sota_x_continued}
\scriptsize
\renewcommand{\arraystretch}{0.68}
\resizebox{\linewidth}{!}{%
\begin{tabular}{lrrrrrrrrrrr}
\hline
\multirow{2}{*}{Instance} & \multicolumn{1}{c}{\multirow{2}{*}{$N_C$}} & \multicolumn{1}{c}{\multirow{2}{*}{BKS}} & \multicolumn{1}{c}{\multirow{2}{*}{$t$}} & \multicolumn{2}{c}{HGS-2012} & \multicolumn{2}{c}{HGS-CVRP} & \multicolumn{2}{c}{MA-NTGA} & \multicolumn{2}{c}{MA-ETGA} \\
\cmidrule(l){5-6} \cmidrule(l){7-8} \cmidrule(l){9-10} \cmidrule(l){11-12}
    & \multicolumn{1}{c}{} & \multicolumn{1}{c}{} & \multicolumn{1}{c}{} & \multicolumn{1}{c}{$f^*$} & \multicolumn{1}{c}{Gap(\%)} & \multicolumn{1}{c}{$f^*$} & \multicolumn{1}{c}{Gap(\%)} & \multicolumn{1}{c}{$f^*$} & \multicolumn{1}{c}{Gap(\%)} & \multicolumn{1}{c}{$f^*$} & \multicolumn{1}{c}{Gap(\%)} \\
\hline
X-n502-k39 & 501 & 69226.0 & 1202.4 & 69264.0 & 0.05 & 69230.0 & 0.01 & 69566.0 & 0.49 & 69564.0 & 0.49 \\
X-n513-k21 & 512 & 24201.0 & 1228.8 & 24201.0 & 0.00 & 24201.0 & 0.00 & 24429.0 & 0.94 & 24509.0 & 1.27 \\
X-n524-k153 & 523 & 154593.0 & 1255.2 & 154713.0 & 0.08 & 154646.0 & 0.03 & 154764.0 & 0.11 & 155331.0 & 0.48 \\
X-n536-k96 & 535 & 94846.0 & 1284.0 & 95125.0 & 0.29 & 95040.0 & 0.20 & 95622.0 & 0.82 & 96355.0 & 1.59 \\
X-n548-k50 & 547 & 86700.0 & 1312.8 & 86888.0 & 0.22 & 86710.0 & 0.01 & 87426.0 & 0.84 & 87699.0 & 1.15 \\
X-n561-k42 & 560 & 42717.0 & 1344.0 & 42738.0 & 0.05 & 42726.0 & 0.02 & 43317.0 & 1.40 & 43388.0 & 1.57 \\
\hline
\end{tabular}%
}
\end{table}

%% file: tables/sota_hg.tex
\begin{table}[!ht]
\centering
\caption{Comparative results on the 90 \textit{GH} benchmark instances for DIMACS, HGS-DIMACS, MA-NTGA, and MA-ETGA.}
\label{tab:sota_hg}
\scriptsize
\renewcommand{\arraystretch}{0.68}
\resizebox{\linewidth}{!}{%
\begin{tabular}{lrrrrrrrrrrr}
\hline
\multirow{2}{*}{Instance} & \multicolumn{1}{c}{\multirow{2}{*}{$N_C$}} & \multicolumn{1}{c}{\multirow{2}{*}{BKS}} & \multicolumn{1}{c}{\multirow{2}{*}{$t$}} & \multicolumn{2}{c}{DIMACS} & \multicolumn{2}{c}{HGS-DIMACS} & \multicolumn{2}{c}{MA-NTGA} & \multicolumn{2}{c}{MA-ETGA} \\
\cmidrule(l){5-6} \cmidrule(l){7-8} \cmidrule(l){9-10} \cmidrule(l){11-12}
	& \multicolumn{1}{c}{} & \multicolumn{1}{c}{} & \multicolumn{1}{c}{} & \multicolumn{1}{c}{$f^*$} & \multicolumn{1}{c}{Gap(\%)} & \multicolumn{1}{c}{$f^*$} & \multicolumn{1}{c}{Gap(\%)} & \multicolumn{1}{c}{$f^*$} & \multicolumn{1}{c}{Gap(\%)} & \multicolumn{1}{c}{$f^*$} & \multicolumn{1}{c}{Gap(\%)} \\
\hline
C1\_2\_1 & 200 & 2698.6 & 1800.0 & 2698.6 & 0.00 & 2698.6 & 0.00 & 2698.6 & 0.00 & 2698.6 & 0.00 \\
C1\_2\_5 & 200 & 2694.9 & 1800.0 & 2694.9 & 0.00 & 2694.9 & 0.00 & 2694.9 & 0.00 & 2694.9 & 0.00 \\
C1\_2\_10 & 200 & 2624.7 & 1800.0 & 2624.7 & 0.00 & 2624.7 & 0.00 & 2624.7 & 0.00 & 2624.7 & 0.00 \\
C2\_2\_1 & 200 & 1922.1 & 1800.0 & 1922.1 & 0.00 & 1922.1 & 0.00 & 1922.1 & 0.00 & 1922.1 & 0.00 \\
C2\_2\_5 & 200 & 1869.6 & 1800.0 & 1869.6 & 0.00 & 1869.6 & 0.00 & 1869.6 & 0.00 & 1869.6 & 0.00 \\
C2\_2\_10 & 200 & 1791.2 & 1800.0 & 1791.2 & 0.00 & 1791.2 & 0.00 & 1791.2 & 0.00 & 1791.2 & 0.00 \\
\hline
\end{tabular}%
}
\end{table}

\begin{table}[!ht]
\centering
\caption{Comparative results on the 90 \textit{GH} benchmark instances for DIMACS, HGS-DIMACS, MA-NTGA, and MA-ETGA (continued).}
\label{tab:sota_hg_continued}
\scriptsize
\renewcommand{\arraystretch}{0.68}
\resizebox{\linewidth}{!}{%
\begin{tabular}{lrrrrrrrrrrr}
\hline
\multirow{2}{*}{Instance} & \multicolumn{1}{c}{\multirow{2}{*}{$N_C$}} & \multicolumn{1}{c}{\multirow{2}{*}{BKS}} & \multicolumn{1}{c}{\multirow{2}{*}{$t$}} & \multicolumn{2}{c}{DIMACS} & \multicolumn{2}{c}{HGS-DIMACS} & \multicolumn{2}{c}{MA-NTGA} & \multicolumn{2}{c}{MA-ETGA} \\
\cmidrule(l){5-6} \cmidrule(l){7-8} \cmidrule(l){9-10} \cmidrule(l){11-12}
	& \multicolumn{1}{c}{} & \multicolumn{1}{c}{} & \multicolumn{1}{c}{} & \multicolumn{1}{c}{$f^*$} & \multicolumn{1}{c}{Gap(\%)} & \multicolumn{1}{c}{$f^*$} & \multicolumn{1}{c}{Gap(\%)} & \multicolumn{1}{c}{$f^*$} & \multicolumn{1}{c}{Gap(\%)} & \multicolumn{1}{c}{$f^*$} & \multicolumn{1}{c}{Gap(\%)} \\
\hline
R1\_4\_1 & 400 & 10305.8 & 3600.0 & 10305.8 & 0.00 & 10305.8 & 0.00 & 10313.8 & 0.08 & 10337.6 & 0.31 \\
R1\_4\_5 & 400 & 9184.6 & 3600.0 & 9188.6 & 0.04 & 9190.3 & 0.06 & 9208.7 & 0.26 & 9198.4 & 0.15 \\
R1\_4\_10 & 400 & 8077.8 & 3600.0 & 8086.4 & 0.11 & 8081.9 & 0.05 & 8109.5 & 0.39 & 8138.7 & 0.75 \\
R2\_4\_1 & 400 & 7520.7 & 3600.0 & 7520.7 & 0.00 & 7520.7 & 0.00 & 7559.7 & 0.52 & 7526.8 & 0.08 \\
R2\_4\_5 & 400 & 6567.9 & 3600.0 & 6567.9 & 0.00 & 6567.9 & 0.00 & 6568.1 & 0.00 & 6577.1 & 0.14 \\
R2\_4\_10 & 400 & 5638.1 & 3600.0 & 5645.9 & 0.14 & 5657.7 & 0.35 & 5690.2 & 0.92 & 5694.0 & 0.99 \\
\hline
\end{tabular}%
}
\end{table}

%% file: tables/sota_jd.tex
\begin{table}[!ht]
\centering
\caption{Comparative results on the 20 \textit{JD} benchmark instances for MA-FIRD, MA-NTGA, and MA-ETGA.}
\label{tab:sota_jd}
\scriptsize
\resizebox{\linewidth}{!}{%
\begin{tabular}{lrrrrrrrrrrr}
\hline
\multirow{2}{*}{Instance} & \multicolumn{1}{c}{\multirow{2}{*}{$N_C$}} & \multicolumn{1}{c}{\multirow{2}{*}{BKS}} & \multicolumn{3}{c}{MA-FIRD} & \multicolumn{3}{c}{MA-NTGA} & \multicolumn{3}{c}{MA-ETGA} \\
\cmidrule(l){4-6} \cmidrule(l){7-9} \cmidrule(l){10-12}
	& \multicolumn{1}{c}{} & \multicolumn{1}{c}{} & \multicolumn{1}{c}{$f^*$} & \multicolumn{1}{c}{$t$} & \multicolumn{1}{c}{Gap(\%)} & \multicolumn{1}{c}{$f^*$} & \multicolumn{1}{c}{$t$} & \multicolumn{1}{c}{Gap(\%)} & \multicolumn{1}{c}{$f^*$} & \multicolumn{1}{c}{$t$} & \multicolumn{1}{c}{Gap(\%)} \\
\hline
F201 & 200 & 64083.4 & 64083.4 & 1360.4 & 0.00 & 64297.8 & 424.0 & 0.33 & 64483.1 & 516.1 & 0.62 \\
F202 & 200 & 64378.1 & 64378.1 & 1236.1 & 0.00 & 64758.0 & 415.1 & 0.59 & 64698.3 & 608.1 & 0.50 \\
F203 & 200 & 65547.4 & 65547.4 & 1349.0 & 0.00 & 65526.9 & 495.1 & -0.03 & 65664.1 & 614.9 & 0.18 \\
F204 & 200 & 64037.3 & 64037.3 & 1429.9 & 0.00 & 64179.1 & 532.0 & 0.22 & 64310.7 & 641.4 & 0.43 \\
F401 & 400 & 115876.0 & 115876.0 & 7005.8 & 0.00 & 116999.0 & 841.8 & 0.97 & 117057.0 & 1056.2 & 1.02 \\
F402 & 400 & 121997.0 & 121997.0 & 7082.8 & 0.00 & 122429.0 & 865.5 & 0.35 & 122369.0 & 1567.1 & 0.30 \\
\hline
\end{tabular}%
}
\end{table}